\documentclass[fleqn,usenatbib]{mnras}

\usepackage{newtxtext,newtxmath}

\usepackage[T1]{fontenc}
\usepackage{natbib}
\usepackage{xspace}
\usepackage{booktabs}
\usepackage[figuresright]{rotating}
\usepackage{pdflscape}
\usepackage{afterpage}

\DeclareRobustCommand{\VAN}[3]{#2}
\let\VANthebibliography\thebibliography
\def\thebibliography{\DeclareRobustCommand{\VAN}[3]{##3}\VANthebibliography}

\usepackage{graphicx}	
\usepackage{amsmath}	

\defcitealias{Ramaiya2025}{R25}
\newcommand{\Ramaiya}{\citetalias{Ramaiya2025}\xspace}

\newcommand{\rev}[1]{#1}
\newcommand{\revtwo}[1]{#1}

\title[TITAN Host Galaxies I]{TITAN DR1: The Type Ia Supernova Trove from ATLAS in the Nearby Universe -- Host Galaxies I: Associations, Redshifts and Derived Properties}

\author[J. W. Tweddle et al.]{
Jack W. Tweddle,$^{1}$\thanks{E-mail: jack.tweddle@physics.ox.ac.uk (JWT)}
Yukei S. Murakami,$^{2}$
Maria Vincenzi,$^{1}$
Stephen J. Smartt,$^{1}$
Bastien Carreres,$^{3}$
\newauthor Fiorenzo Stoppa,$^{1}$
Erik R. Peterson,$^{4,5}$
Brodie Popovic,$^{6}$
Jasper Milstein,$^{7}$
MaKenzie Elliott,$^{7}$
\newauthor Elijah G. Marlin,$^{2,7}$
Dillon Brout,$^{7}$
David O. Jones,$^{8}$
Mitchell Dixon,$^{8}$
Adam G. Riess,$^{2,9}$
\newauthor Daniel M. Scolnic,$^{3}$
and Alejandro Clocchiatti$^{10}$
\\
$^{1}$Astrophysics sub-Department, Department of Physics, University of Oxford, Keble Road, Oxford, OX1 3RH, UK\\
$^{2}$Department of Physics and Astronomy, Johns Hopkins University, Baltimore, MD 21218, USA\\
$^{3}$Department of Physics, Duke University, Durham, NC 27708, USA\\
$^4$Department of Physics, University of Michigan, Ann Arbor, MI 48109, USA\\
$^5$Society of Fellows, University of Michigan, Ann Arbor, MI 48109, USA\\
$^6$School of Physics and Astronomy, University of Southampton, Southampton SO17 1BJ, UK\\
$^7$Departments of Astronomy and Physics, Boston University, Boston MA 02215\\
$^8$Institute for Astronomy, University of Hawai'i, 640 N.~A'ohoku Pl., Hilo, HI 96720, USA\\
$^9$Space Telescope Science Institute, 3700 San Martin Drive, Baltimore, MD 21218, USA\\
$^{10}$Instituto de Astrofisica, Pontificia Universidad Católica de Chile, Vicuña Mackenna 4860, Macul, Santiago, Chile
}

\date{Accepted 2026 November 7. Received 2026 November 4; in original form 2026 July 9}

\pubyear{\the\year{}}

\begin{document}
\label{firstpage}
\pagerange{\pageref{firstpage}--\pageref{lastpage}}
\maketitle

\begin{abstract}
We present the host-galaxy catalogue for TITAN DR1, a sample of 8378 spectroscopically confirmed Type Ia supernovae (SNe~Ia) at $z \lesssim 0.15$ -- the largest homogeneous low-redshift SN Ia sample assembled to date. We describe the association of each SN with its host, the compilation of redshifts, measurements of uniform far-ultraviolet to mid-infrared (FUV--MIR) photometry, and the derivation of host properties through spectral energy distribution (SED) fitting. Combining four host-association methods with manual vetting, we securely identify hosts for 7856 SNe~Ia (93.8\%). We compile host-galaxy spectroscopic redshifts for 5522 objects and provide CMB-frame and peculiar-velocity-corrected redshifts. \rev{PSF-matched} photometry is measured for 7568 host galaxies, including both global fluxes and local optical-only fluxes within a fixed 1.5~kpc aperture centred on the SN; \texttt{Bagpipes} returns successful global fits for \rev{all} hosts and local fits for 7329. Stellar mass and rest-frame $g-z$ colour are robustly measured from optical data alone, whereas reliable estimates of star-formation rate, dust attenuation, and mass-weighted age require broader wavelength coverage to break age--dust degeneracies\rev{, and are reported for 5827 hosts}; in our redshift range, MIR data suffice and far-IR coverage is not required. The hosts span a representative low-redshift galaxy population, with a median stellar mass of $\log_{10}(M_*/M_\odot)=10.22$ (\rev{60\%} high-mass; $\log_{10}(M_*/M_\odot)>10$), \rev{71\%} star-forming ($\log_{10}(\mathrm{sSFR}/\mathrm{yr}^{-1}) > -10.5$), and an approximately exponential $A_V$ distribution. This catalogue provides the foundation for forthcoming analyses of the luminosity step and the broader role of galaxy environment in SN~Ia standardization and cosmology.
\end{abstract}




\section{Introduction}\label{sec:intro}

Type Ia supernovae (SNe~Ia) are among the most powerful probes of the expansion history of the Universe. The empirical relationship between light curve shape and luminosity established by \citet{Phillips1993}, together with later multicolour and two-parameter standardization formalisms \citep{Riess1996, Tripp1998}, made it possible to use SNe~Ia as precision distance indicators. Observations of distant SNe~Ia then revealed that the cosmic expansion is accelerating, providing the first direct evidence for dark energy or a cosmological constant \citep{Riess1998, Perlmutter1999}. Since then, successive improvements in light curve modelling, calibration, bias corrections, and sample size have transformed SN~Ia cosmology into a precision field, with analyses based on increasingly large compilations such as SNLS, JLA, Pantheon, Pantheon+, \revtwo{DES5YR and, most recently, Unite} \citep{Guy2007, Guy2010, Betoule2014, Scolnic2018, Brout2022, Sanchez2024, Camilleri2026, Lee2026}.

The success of SN~Ia cosmology rests on the fact that SNe~Ia can be standardized. In modern analyses, this is usually done by fitting the SN light curves with models such as SALT and correcting the inferred luminosity using at least light curve width and colour \citep{Guy2007, Kenworthy2021}. Even after these corrections, however, Hubble diagram residuals retain measurable correlations with the properties of the galaxies in which the SNe explode. The best known of these is the host-galaxy stellar-mass step, in which SNe~Ia in more massive galaxies are systematically brighter in standardized luminosity than those in less massive hosts \citep{Sullivan2010,Kelly2010, Lampeitl2010}. Related trends have also been reported with global and local star formation, metallicity, colour, stellar population age, and galactocentric environment \citep[e.g.,][]{DAndrea2011, Gupta2011, Childress2013, Rigault2013, Uddin2017, Rigault2020, Kelsey2021, Kelsey2023, Toy2025}. These correlations are now a central part of the astrophysical and systematic-error landscape of SN~Ia cosmology.

Understanding these host-galaxy correlations is important for two reasons. First, they may encode real information about SN~Ia progenitor channels and explosion physics. Host stellar mass is unlikely to be a fundamental parameter; rather, it is expected to reflect a network of more directly relevant quantities, including stellar age, recent star formation, metallicity, and dust content \citep{Sullivan2010,Childress2013}. Recent work has sharpened the question of whether some of these quantities, especially stellar population age or environmentally-dependent dust properties, may be more physically informative than stellar mass itself \citep[e.g.,][]{BroutScolnic2021, Lee2022, Briday2022, Wiseman2023, Popovic2023, Larison2024}. Second, because host-galaxy corrections are applied directly in cosmological analyses, uncertainty in their origin, measurement, and redshift evolution propagates into the final error budget on cosmological parameters \citep[e.g.,][]{Brout2022, Vincenzi2024, Popovic2026}. Building large, homogeneous samples with well-characterized host galaxies is therefore essential both for improving SN standardization and for understanding the astrophysics that underlies it.

This motivates a renewed focus on the nearby-Universe SN~Ia population. Low-redshift samples play a special role in cosmology: they anchor the Hubble diagram \citep[e.g.,][]{Foley2018, Riess2022, Rigault2025, Sherman2026}, permit the most precise host-galaxy redshifts \citep[e.g.,][]{Carr2022, Peterson2022}, and provide access to host galaxies that are sufficiently bright and well resolved to enable detailed environmental studies \citep[e.g.,][]{Hamuy2000, Sullivan2010, Rigault2013, Roman2018, Galbany2018,Smith2020}. In this regime, global galaxy photometry can be measured at high signal-to-noise across much of the sky, and local measurements near the SN position can often be resolved on kiloparsec scales \citep[e.g.,][]{Kelsey2021, Kelsey2023, Ginolin2025_colour, ginolinZTFDR2_2025, Kelsey2026}. At the same time, the low-redshift Universe is where peculiar velocities and host-identification ambiguities \citep[e.g.,][]{Qu2024} matter most, so constructing a clean and internally consistent host catalogue requires careful treatment of host association, redshift compilation, and photometry.

Built from SNe Ia observed by the Asteroid Terrestrial-impact Last Alert System (ATLAS), the \textit{Type Ia Supernova Trove from ATLAS in the Nearby Universe} \citep[TITAN; Murakami et al., in prep.;][]{Marlin2025} is designed to serve as the premier low-redshift SN~Ia sample for cosmology in the era of Stage~IV dark energy experiments. Spanning $z \lesssim 0.15$, TITAN comprises 8378 spectroscopically confirmed SNe~Ia -- the largest homogeneous nearby SN~Ia sample assembled to date. It therefore combines two features that are rarely available simultaneously: low redshift and large statistical power. 

\begin{figure*}
    \centering
    \includegraphics[width=\textwidth]{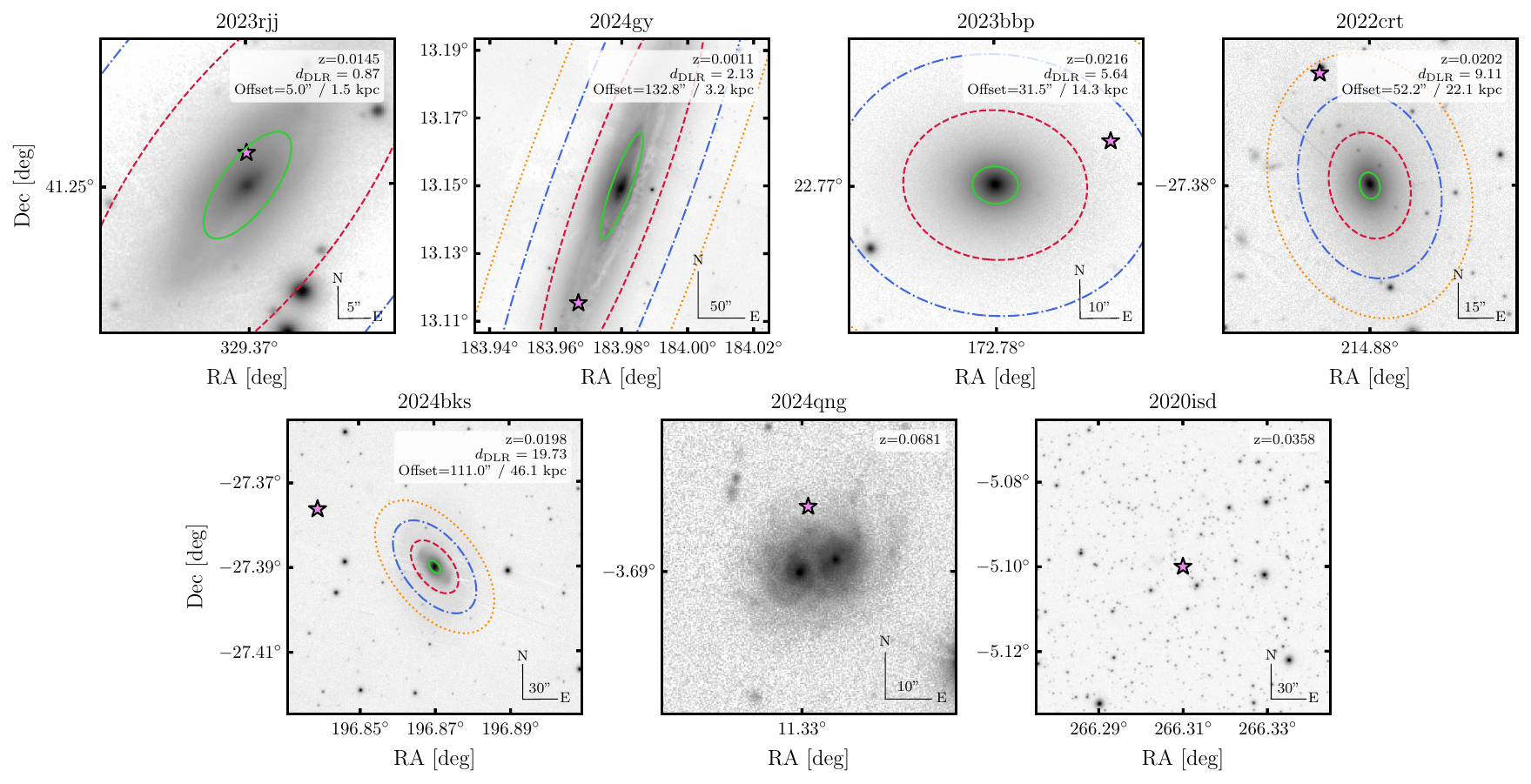}
    \caption{A selection of representative host-galaxy association cases from the TITAN sample, shown using $i$-band Pan-STARRS cutouts. In all panels, the location of the supernova is marked by a pink star. Where applicable, elliptical contours indicate constant $d_\mathrm{DLR}$ values of 1 (green, solid), 4 (red, dashed), 7 (blue, dash-dotted), and 10 (orange, dotted). \textit{Top:} From left to right, an example of a successful host association with $d_\mathrm{DLR}\leq1$, $1<d_\mathrm{DLR}\leq4$, $4<d_\mathrm{DLR}\leq7$, $7<d_\mathrm{DLR}\leq10$, respectively. \textit{Bottom:} Examples of failed host associations. Left: a candidate host is identified, but $d_\mathrm{DLR}$ exceeds the association threshold of 12. Middle: the supernova lies at the intersection of two merging galaxies, rendering the host association ambiguous. Right: no plausible host galaxy is detected near the supernova, and the event is classified as hostless.}
    \label{fig:association_examples}
\end{figure*}

For large samples of SNe Ia like TITAN, photometry-based spectral energy distribution (SED) fitting offers a practical route to uniform host-galaxy characterization, with recently developed pipelines now able to perform host association, photometry, and SED fitting at scale in response to transient discovery (e.g., the \texttt{Blast}/\texttt{FrankenBlast} framework; \citealt{Jones2024, Nugent2026}). In practice, such characterization typically relies on rest-frame optical photometry alone, which is generally sufficient to recover stellar masses and rest-frame colours with high precision owing to the tight relationship between colour and mass-to-light ratio \citep{Taylor2011}. Optical-only host masses can therefore be measured relatively uniformly from low to high redshift, and have become the standard host-galaxy observable used in SN~Ia cosmology analyses \citep[e.g.,][]{Wiseman2020,Smith2020}.  An analysis of optical host galaxy data was recently employed in the Zwicky Transient Facility (ZTF) SN Ia DR2 analysis \citep{Ginolin2025_colour, ginolinZTFDR2_2025} to probe the effect of host environment on the standardization of low redshift SNe. 

However, quantities such as recent star-formation rate, dust attenuation, and mass-weighted age are more difficult to infer from optical data alone because of the well-known degeneracies between stellar age, dust, and metallicity \citep{Worthey1994, Papovich2001}. Breaking these degeneracies generally requires broader wavelength coverage, with optical data augmented by rest-frame ultraviolet (UV) and near- (NIR), mid- (MIR), and (in some cases) far-infrared (FIR) photometry. UV fluxes are sensitive to recent, unobscured star formation, while NIR anchors the older, redder stellar continuum and can improve mass estimates for evolved stellar populations. The MIR traces dust-reprocessed emission, including polycyclic aromatic hydrocarbon (PAH) features at $\sim 6-12~\mu\text{m}$, and can capture the initial rise of the dust-emission curve in low-redshift galaxies. \rev{Broader wavelength coverage does not, however, break the degeneracy between stellar age and metallicity; robust metallicity measurements instead require spectroscopy, either through metallicity-sensitive absorption features or full spectral fitting \citep[e.g.,][]{Gallazzi2005, Conroy2013}. In our SED fits, stellar metallicity is included as a free parameter (Table~\ref{tab:bagpipes_params}) and marginalised over, so that its uncertainty is propagated into the other derived properties, but the resulting metallicity estimates are prior-dominated for most hosts, and we therefore do not include them in the TITAN host catalogue.}

Recent host-galaxy studies have highlighted the importance of this broader wavelength coverage. In \citet[hereafter \Ramaiya]{Ramaiya2025}, the DES5YR host-galaxy sample was analysed using optical--FIR photometry, and it was shown that FIR data are crucial for constraining the dust-emission budget and for mitigating age--dust--metallicity degeneracies at intermediate redshift ($z\lesssim0.6$). However, FIR observations of the required sensitivity are typically limited to relatively small regions of the sky observed by facilities such as \textit{Herschel}, and are therefore not readily available for the all-sky or wide-field SN samples that underpin modern cosmological analyses. Establishing whether FIR data are genuinely required for robust host-galaxy characterization is therefore of considerable practical importance.

For the much lower redshift galaxies targeted by TITAN, it is plausible that MIR measurements (such as those from the \textit{Wide-field Infrared Survey Explorer} (\textit{WISE}; \citealt{Wright2010}), covering the $3.4$--$22~\mu$m range) may already capture sufficient information from the dust-emission curve to recover comparably robust constraints without requiring FIR observations for every object. This question is particularly relevant for TITAN, because the availability of all-sky archival imaging from the far-ultraviolet (FUV) through the MIR makes it feasible to assemble homogeneous UV--MIR SEDs for a very large sample of nearby SN~Ia host galaxies.

In this paper, we present the host-galaxy foundation for the TITAN DR1 SN~Ia sample. In Section~\ref{sec:host_association} we describe the procedure used to associate each SN with a host galaxy. In Section~\ref{sec:redshifts} we compile spectroscopic and SN-based redshifts and describe the corresponding peculiar-velocity corrections. In Section~\ref{sec:photometry} we present the multi-band photometry, spanning the FUV to the MIR. In Section~\ref{sec:sed_fitting} we model these data to derive host-galaxy properties and assess which parameters can be robustly recovered with different wavelength coverage. Finally, in Section~\ref{sec:sample_props} we present the derived global and local host-galaxy properties and characterize the demographic properties of the TITAN host population. A companion analysis will use these measurements to investigate the host-galaxy luminosity step and the broader role of galaxy environment in SN~Ia standardization. Where relevant throughout this work, we assume a \cite{Planck2020} cosmology, with $H_0=67.66\ \mathrm{km\ s^{-1}\ Mpc^{-1}}$ and $\Omega_{M,0}=0.30966$.

\section{The TITAN Sample and Host Galaxy Association}\label{sec:host_association}

As detailed in Murakami et al. (\textit{in prep.}), the TITAN sample is constructed from the 10,409 spectroscopically classified SNe~Ia reported to the Transient Name Server (TNS)\footnote[2]{\url{https://www.wis-tns.org/.}} between 1 January 2016 and 7 February 2025, for which the ATLAS Forced Photometry Server \citep{Shingles2021} returns at least one observation. Following reclassification with \texttt{SNID-SAGE} \citep{SNID-SAGE}, 8378 objects retain a confident normal SN~Ia classification and comprise the TITAN bronze sample. Imposing the additional requirement of a cosmology-grade light curve, defined as a well-sampled light curve yielding a well-constrained SALT \citep{Guy2007, Kenworthy2021} fit, reduces the sample to $\sim4500$ SNe~Ia; these constitute the TITAN silver sample. Finally, further requiring a spectroscopic host-galaxy redshift (Section~\ref{sec:redshifts}) yields $\sim3000$ SNe~Ia, which form the TITAN gold sample and are the subset best suited to cosmological analyses. In this section, we describe the procedure used to associate each SN in the bronze sample with its host galaxy.

\subsection{Initial association}

The identification of host galaxies of transients is often performed \citep[e.g.,][]{Sako2018, Smith2020} using the directional light radius (DLR) method \citep{Sullivan2006,Gupta2016}. In brief, the method assigns a host by normalizing the supernova-galaxy angular separation ($d_{\perp}$) by the galaxy’s projected light radius in the direction of the supernova ($R_\mathrm{DLR}$). For each candidate galaxy, one measures its shape (semi-major axis $a$, semi-minor axis $b$, and position angle $\theta$) from imaging, and defines the corresponding light radius of the galaxy in the direction of the SN. This is often taken to be the ellipse defined by the second-order moment of the galaxy's light distribution \citep[e.g.,][]{Sullivan2006, Sako2018}, but sometimes uses the parameters of the aperture used for photometry (Smith et al., \textit{in prep.}). The normalized separation:
\begin{equation}
    d_{\mathrm{DLR}} \equiv \frac{d_{\perp}}{R_{\mathrm{DLR}}} \label{eq:ddlr}
\end{equation}
compares how far the SN lies relative to the galaxy’s own extent and orientation rather than in absolute distance. The host is then often taken to be the galaxy with the smallest $d_{\mathrm{DLR}}$, subject to a survey-dependent maximum cut to flag `hostless' cases; ties can be broken with secondary information (e.g., brightness, redshift consistency). Representative examples of the host-galaxy association outcomes in TITAN are shown in Figure \ref{fig:association_examples}.

To maximize the information available when selecting candidate host galaxies, we combine information from four host association methods: 
\begin{enumerate}
    \item The ATLAS100 sample \citep{Srivastav2026}, whose SNe Ia are a subset of the TITAN sample. ATLAS100 is a volume-limited sample of all astrophysical transients within 100~Mpc observed by ATLAS between 2017 and 2023. All objects in the sample have been manually vetted -- including host assignments -- so for any SN present in the ATLAS100 sample we initially adopt the ATLAS100 host without modification.
    
    \item The probabilistic host-galaxy association code, \texttt{Pröst}\footnote[3]{\url{https://github.com/alexandergagliano/Prost.}} \citep{Prost}. \texttt{Pröst} computes a Bayesian posterior probability for each potential host by combining information such as the $d_\text{DLR}$, redshift, and galaxy brightness, and returns the most probable host within a given search cone centred on the transient. We define the following search parameters: a search cone of $60^{\prime\prime}$; a uniform redshift prior between $0.0001 \leq z \leq 0.25$; a uniform offset prior between $0 \leq d_\text{DLR} \leq 15$ with a $\Gamma$ function likelihood that disfavours hosts with larger fractional offsets; a wide uniform prior on host absolute magnitude of $-30 \leq M_{\text{host}} \leq 20$ with a likelihood favouring more luminous hosts, $\ln \mathscr{L} \propto -M_{\text{host}}$, reflecting the empirical scaling of the supernova rate with host luminosity \citep{Li2011}. This method searches the GLADE+ \citep{Glade+}, the 9$^{\text{th}}$ and 10$^{\text{th}}$ data releases of the Dark Energy Camera Legacy Survey \citep[DECaLS DR9 \& DR10;][]{LegacySurvey}, Pan-STARRS DR1 and DR2 \citep{PanSTARRS}, and SkyMapper DR4 \citep{SkyMapper} catalogues. We search all available surveys for each transient, with the galaxy with the highest probability being selected as the best-matching host from \texttt{Pröst}.
    
    \item $d_\mathrm{DLR}$ matching within the DESI Legacy Survey \citep{LegacySurvey}. The top three candidate hosts are returned, ranked by their $d_\mathrm{DLR}$ likelihood. Any candidates that are not labelled as having a \texttt{GALAXY} spectroscopic type and those with a spectroscopic redshift $z \geq 0.25$ or a $d_{\mathrm{DLR}}>15$ are automatically discarded.
    
    \item \texttt{Sherlock}\footnote[4]{\url{https://github.com/thespacedoctor/sherlock}} \citep{Sherlock} -- a boosted decision-tree algorithm that crossmatches transient positions against a large, locally stored library of all-sky and specialist catalogues. These catalogues include many of those listed above, as well as the 12$^{\text{th}}$ data release of the Sloan Digital Sky Survey \citep[SDSS DR12;][]{SDSS_DR12}, the Two Micron All-Sky Survey \citep[2MASS;][]{2MASS}, the Local AGN Survey Galaxy catalogue \citep[LASr;][]{LASR} and the NASA Extragalactic Database Redshift-Independent Distance catalogue \citep[NED-D;][]{NED-D}. \texttt{Sherlock} performs parallelized positional and morphological crossmatching across these catalogues to identify potential host galaxies and a contextual classification of the transient. Candidate host galaxies are rejected if the transient lies beyond 1.2 times the catalogued semi-major axis of the galaxy,\footnote[5]{Note that this will differ from the ellipse defined by the second-order moment of the light profile.} or if a distance estimate places the transient more than $\sim 50$~kpc from the core.
\end{enumerate}

\begin{figure}
\centering
\includegraphics[width=\columnwidth]{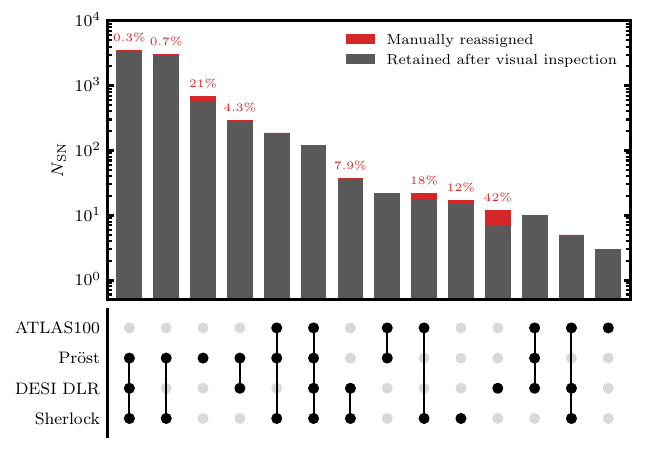}
\caption{\rev{Agreement between the four initial host-association methods for the TITAN SN Ia sample. Each column corresponds to a combination of methods (indicated by the connected dots below) whose candidate host coincides with the adopted host, and the bar shows the number of SNe for which that combination supports the adopted host. The red portion of each bar shows associations subsequently reassigned during visual inspection, noting a logarithmic $y$-axis, and numbers above the bars give the percentage of associations in each combination that were reassigned.}}
\label{fig:host_agreement}
\end{figure}

Across the four methods we split candidate hosts into groups whose reported galaxy-core coordinates agree within $2^{\prime\prime}$ (to allow for catalogue-to-catalogue astrometric differences). The adopted host is taken from the group with the largest membership (majority agreement). If multiple groups have the same size, we break ties by (i) preferring the group that includes the ATLAS100 match (if available), (ii) preferring the group containing the host with the smallest $d_\mathrm{DLR}$ (if available), and, if still tied, (iii) preferring the group containing the host with the smallest angular offset from the SN. At least one candidate host galaxy is initially identified for 8284 (98.9\%) of TITAN SNe Ia. 

\rev{The four methods show a high degree of consistency: for 90.9\% of SNe, the adopted host is supported by at least two methods (43.3\% by two, 46.1\% by three and 1.5\% by all four), with the full distribution of agreeing combinations shown in Figure~\ref{fig:host_agreement}. The methods differ substantially in coverage. \texttt{Pröst} and \texttt{Sherlock} operate across the full sky and support the adopted host in 98.8 and 87.0\% of cases, respectively; the DESI DLR method is restricted by construction to galaxies with DESI spectra, for which it supports the adopted host in 92.6\% of cases. The manually vetted ATLAS100 subsample provides an independent benchmark, against which \texttt{Pröst}, \texttt{Sherlock} and the DESI DLR method (where available) recover the ATLAS100 host in 91.9, 90.5 and 92.4\% of cases, respectively. \texttt{Pröst} is the dominant contributor, being the sole method supporting the adopted host for 697 SNe (8.7\%), whereas each of the other methods is the sole supporter for fewer than 20. The number of agreeing methods is a strong indicator of reliability: 75\% of the associations reassigned during visual inspection (Section~\ref{sec:visual_inspection}) were supported by only a single method, corresponding to reassignment rates of 21.3, 1.2, 0.2 and 0\% for associations supported by one, two, three and four methods, respectively. For the much larger samples expected from surveys such as LSST, where inspecting every association by eye will be infeasible, we therefore recommend combining several independent host-association methods as standard practice.}

\subsection{Visual inspection} \label{sec:visual_inspection}

During the process of obtaining photometry for each candidate galaxy (see Section \ref{sec:photometry} for details), we visually inspected \rev{optical co-add images (Section~\ref{sec:masking})} to verify the host association. This assessment was supported by a host-association confidence score computed for every transient. The score combined several metrics: DLR separation, the agreement between the catalogued galaxy spectroscopic redshift, $z_\mathrm{host,spec}$ and the SN spectroscopic redshift, $z_\mathrm{SN}$ (see Section \ref{sec:redshifts} for details on the compilation of redshifts in the TITAN sample), and the \texttt{Pröst} association probability. Consistent, high-quality measurements increased the confidence score, whereas missing or conflicting information incurred penalties. Systems with low confidence scores or with $|z_\mathrm{host,spec} - z_\mathrm{SN}| \geq 0.01$ were explicitly flagged as potentially problematic during visual inspection. This process led to the manual reassignment of the most likely host galaxy in 250 cases (3.0\%). A further 165 SNe were found to have no viable candidate hosts upon further inspection (due to, e.g., the initially identified candidate galaxy being found to be a point source such as a star, and/or visual inspection of a 100 kpc radius around the SN location failing to identify any viable candidate hosts), bringing the `true'\footnote[6]{In this instance, `true' hostless transients are defined as those that do not have any identified candidate hosts; transients with an identified host galaxy with $d_\mathrm{DLR}>12$ are not included in this fraction.} hostless fraction to 3.1\% (259 SNe\footnote[7]{In this work, we do not attempt to infer host-galaxy properties for these orphan SNe, and instead reserve a dedicated analysis of this population within TITAN for future work.}). This is consistent with the fraction of hostless SNe observed in similar low-$z$ samples \citep[e.g., ZTF,][reporting a fraction of $2.2\pm0.6\%$]{Perley2020}.

Additionally, 75 (0.9\%) supernovae are found to have ambiguous host associations upon further inspection. Ambiguities can arise for several reasons, including supernovae occurring in merging systems, events projected near the centres of galaxy clusters with multiple nearby candidates, or cases where multiple viable candidate hosts with similar $d_\mathrm{DLR}$ are present but insufficient redshift information prohibits a definitive assignment.

\subsection{Cutting by $d_\mathrm{DLR}$} \label{sec:dlr_cuts}

Finally, to ensure a consistent method for host association, we compute the $d_\mathrm{DLR}$ for all candidate galaxies. We use the second-order moment ellipse parameters \texttt{A} (semi-major axis), \texttt{B} (semi-minor axis), and \texttt{THETA} (position angle) as measured by \texttt{sep}, a Python implementation of the Source Extractor algorithms \citep{SExtractor, sep}, \rev{on optical co-add images (Section~\ref{sec:masking})}. Unlike model-dependent quantities such as S\'ersic half-light radii, these moment-based parameters provide a model-independent estimate of the observed galaxy shape and orientation.

We take the celestial coordinates of the supernova ($x_\mathrm{SN}, y_\mathrm{SN}$) as reported on the TNS, and the galaxy centroid coordinates ($x_\mathrm{gal}, y_\mathrm{gal}$) as detected by \texttt{sep}. Neglecting the small positional uncertainties in these quantities, the angular-separation between the SN and galaxy centre is
\begin{equation}
   d_\perp = \sqrt{(\delta x \cos \bar{y})^2+\delta y^2}.
\end{equation}
where $\bar{y}=(y_\mathrm{SN}+y_\mathrm{gal})/2$.

The direction from the galaxy centre to the SN is then 
$\phi=\arctan(\delta y/(\delta x \cos \bar{y}))$. 
Because the galaxy ellipse is rotated relative to the sky coordinate axes, we transform into a coordinate frame aligned with the galaxy's principal axes. The angle $\alpha=\phi-\texttt{THETA}$ then represents the direction to the SN in a frame where the galaxy's semi-major and semi-minor axes lie along the $x$- and $y$- axes, respectively. The DLR of the galaxy is then
\begin{equation}
R_\mathrm{DLR}=\frac{\texttt{A}\cdot\texttt{B}}{\sqrt{[\texttt{A}\cdot\sin(\alpha)]^2 + [\texttt{B}\cdot\cos(\alpha)]^2}}, \label{eq:R_DLR}
\end{equation}
and $d_\mathrm{DLR}$ is computed as outlined in Eq. \ref{eq:ddlr}.

\begin{figure}
    \centering
    \includegraphics[width=\columnwidth]{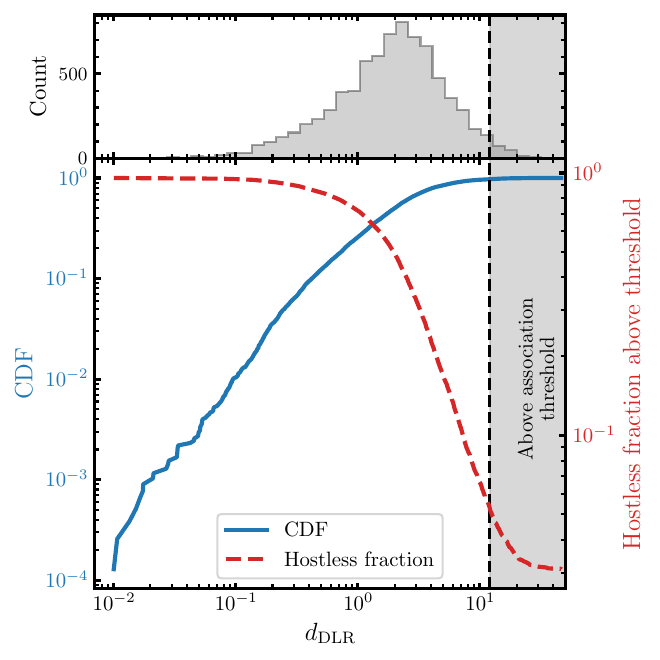}
    \caption{Distribution of $d_\mathrm{DLR}$ in the TITAN sample. In both panels, the vertical dashed line marks the adopted association threshold of $d_\mathrm{DLR}=12$, with the shaded region indicating the regime where SNe in TITAN are no longer associated with a host galaxy. \textit{Top:} Histogram of $d_\mathrm{DLR}$ values. \textit{Bottom:} Cumulative distribution function (CDF; solid blue, left axis) for confidently associated SN~Ia hosts. The dashed red curve (right axis) shows the hostless fraction that would result from adopting the corresponding $d_\mathrm{DLR}$ value as the association threshold; it counts both SNe whose $d_\mathrm{DLR}$ exceeds that threshold and events with no detected host candidates.}
    \label{fig:dlr_cdf}
\end{figure}

The distribution of $d_\mathrm{DLR}$ values for TITAN host-galaxy associations is shown in Figure~\ref{fig:dlr_cdf}. Considering SNe with a confidently identified host, the majority of associations occur at small $d_\mathrm{DLR}$: 25.8\% of SNe have $d_\mathrm{DLR}<1$, 78.9\% have $d_\mathrm{DLR}<4$, 91.9\% have $d_\mathrm{DLR}<7$, and 97.6\% have $d_\mathrm{DLR}<12$. This demonstrates that most supernovae are located well within the characteristic light profiles of their hosts, with only a small tail extending to larger separations.

While some surveys (e.g., DES) adopt a more conservative cutoff of $d_\mathrm{DLR}=4$ for host association, our data suggests that such a threshold is overly restrictive for the low-redshift regime probed by the TITAN sample. In particular, applying a cut at $d_\mathrm{DLR}=4$ would result in an unrealistically large fraction of apparently hostless SNe ($\sim 20\%$). We therefore adopt a more permissive threshold of $d_\mathrm{DLR}=12$, which better balances completeness against the inclusion of spurious associations. Further physical justification for modifying the host association threshold is provided in Appendix~\ref{sec:app_larger_dlr}.

Considering the 188 SNe Ia with $d_\mathrm{DLR}>12$, the 259 SNe Ia with no identified candidate host galaxies and the 75 ambiguous cases, we find that 7856 SNe Ia (93.8\%) in the TITAN sample have a successful host galaxy association. A complete summary of all sample cuts is provided in Table~\ref{tab:reliable_params}.

\begin{table*}
\centering
\caption{Summary of the main sample cuts applied between host association and the fiducial SED-fitting analysis. The `Retained' column gives the number of SNe Ia remaining after each stage, while `Removed' gives the number excluded at that step only. Bold numbers indicate the number of SNe Ia entering that stage of analysis.}
\label{tab:reliable_params}
\begin{tabular*}{\textwidth}{@{\extracolsep{\fill}} l l r r}
\hline
Stage & Requirement / definition & Retained & Removed \\
\hline
\multicolumn{2}{l}{\textbf{General}} & \textbf{8378} & \\
Reliable host association$^{a}$ & Exclude 259 hostless, 75 ambiguous, and 188 with $d_\mathrm{DLR}>12$ & 7856 & 522 \\
Optical imaging coverage & Data for at least one optical survey available & 7728 & 128 \\
\noalign{\smallskip}
\multicolumn{2}{l}{\textbf{Global}} & \textbf{7728} & \\
Masking & Successful field mask construction & 7667 & 61 \\
Global aperture & Reliable Kron aperture measurement & 7568 & 99 \\
Reliable mass / colour sample & Successful \texttt{Bagpipes} fit (reported for all converged fits) & 7568 & 0 \\
\rev{UV and MIR coverage} & \rev{Measurement in both GALEX bands and in W3 or W4} & \rev{5842} & \rev{1726} \\
\rev{Reliable SFR / $A_V$ / $t_{\rm MWA}$ sample} & \rev{No evidence of a luminous AGN: excludes W1$-$W2\,$\geq$\,0.8 with W2\,$<$\,15.05 (Vega)} & \rev{5827} & \rev{15} \\
\noalign{\smallskip}
\multicolumn{2}{l}{\textbf{Local}} & \textbf{7568} & \\
Non-overlapping aperture & Local aperture does not overlap with any masked objects & 7500 & 68 \\
Appropriate aperture size & Local aperture is not larger than global aperture & 7412 & 88 \\
Redshift cut & Require $z\leq0.15$ to report local properties & 7377 & 35 \\
Reliable mass / colour sample & Successful \texttt{Bagpipes} fit (reported for all converged fits) & 7329 & 48 \\
\hline
\end{tabular*}
\vspace{2mm}
\begin{minipage}{\textwidth}
\footnotesize
$^{a}$\,The host-association stage includes 250 manual host reassignments during visual inspection, but these objects remain in the sample and are therefore not counted as removals.
\end{minipage}
\end{table*}

\subsection{Siblings \rev{\& Cousins}}\label{sec:siblings}

A subset of host galaxies in the TITAN sample are `parent' galaxies to more than one SN~Ia. Such SNe are referred to as `siblings' \citep[e.g.,][]{Scolnic2020_siblings, Kelsey2024_siblings}: two or more SNe~Ia that share a common host. Because siblings reside at a common distance and in a shared global environment, they provide a largely model-independent test of SN~Ia standardization -- any residual difference in their standardized distances directly probes the component of the intrinsic scatter that is not attributable to shared host-galaxy properties \citep{Scolnic2020_siblings}.

We identify siblings as a by-product of the host-association procedure described above: any parent galaxy associated with more than one TITAN SN~Ia is flagged as hosting a sibling family. We find 34 parent galaxies hosting a total of 68 sibling SNe~Ia, and reserve a detailed analysis of the TITAN siblings for future work.

\rev{Supernova `cousins' are SNe~Ia that occur in different host galaxies within the same galaxy group or cluster. \revtwo{Galaxies within a group share a common bulk motion \citep{Peterson2022}, and their properties are correlated with their environment at the population level \citep[e.g.,][]{Dressler1980, Helsdon2003, Peng2010}. Cousins therefore offer many of the advantages of siblings from a substantially larger sample.} Using the TITAN host catalogue presented here, \citet{Blum2026} identify 80 cousins in 32 groups at $0.01 < z < 0.05$, and find tentative evidence for $\sim$30--40\% lower distance scatter among cousins than across the full sample.}

\section{Redshifts}\label{sec:redshifts}

In SN cosmology, it is generally preferable to use host-galaxy spectroscopic redshifts rather than those derived from SN spectra, since the latter are non-repeatable measurements and the large, time dependent,  expansion velocities of the supernova ejecta -- typically of order $v\sim 10^4$~km/s \citep[e.g.,][]{Foley2011} -- can introduce degeneracies and uncertainties in the redshift determination that can bias cosmological parameter inference \citep{Steinhardt2020}. In contrast, emission and absorption lines in galaxy spectra are far less affected by velocity broadening ($v \sim 10^2$~km/s), yielding more reliable and precise redshifts -- with typical uncertainties of $\sigma_z \sim 2\times10^{-4}$ compared to $\sigma_z \sim 5\times10^{-3}$ from supernova spectra \citep{Carr2022} -- and thus enabling more accurate cosmological measurements. We here outline different sources of redshifts in the TITAN sample.

\subsection{Host galaxy spectroscopic redshifts}

Upon the successful association of SNe Ia with their host galaxies, we proceed to query a number of publicly available redshift catalogues to compile the most reliable spectroscopic redshift for each system. Where multiple measurements are available, we adopt a hierarchical approach that favours higher-resolution or higher signal-to-noise observations over lower-quality determinations.

In summary, we query the following catalogues in priority order, filtering by candidates with $z\leq0.25$:

\begin{enumerate}
    \item DESI DR1 \citep{DESI_DR1}, the first public data release from the Dark Energy Spectroscopic Instrument, which includes spectra for more than 18 million unique targets from Main Survey observations taken between May 2021 and June 2022, together with Survey Validation data reprocessed with the same reduction pipeline as the Main Survey. We adopt the DR1 quality definition of `useful spectra' (\texttt{ZCAT\_PRIMARY}=True, \texttt{OBJTYPE}=\texttt{`TGT'}, and \texttt{ZWARN}=0), and exclude spectra with \texttt{SPECTYPE}=\texttt{`STAR'}.
    \item MOSTHosts \citep{Soumagnac2024}, a dedicated DESI follow-up programme designed to identify and obtain spectroscopic redshifts for transient host galaxies lacking secure measurements in existing surveys; the catalogue currently contains data for $\sim$41,000 transients.\footnote[8]{The MOSTHosts catalogue does not report redshift uncertainties. However, DESI spectroscopic redshifts typically achieve a precision of $\sim$0.01\%. We therefore adopt this value as a representative uncertainty for MOSTHosts redshifts.}
    \item NED-LVS \citep{NED-LVS}, the NED (NASA/IPAC Extragalactic Database) Local Volume Sample, a curated subset of \(\sim 2 \times 10^6\) NED objects with distances out to 1000~Mpc. The catalogue combines fiducial heliocentric redshifts with redshift-independent distances, with the latter prioritized at distances below 200 Mpc. We use the spectroscopic subset of NED-LVS redshifts, and take the reported $z_\mathrm{unc}$ values as the redshift uncertainties when available; if no $z_\mathrm{unc}$ is available, we adopt a fallback value of 1\%.
    \item NED, via the \texttt{astroquery} interface. Candidate matches are required to have a valid redshift and a spectroscopic classification, identified using the NED \texttt{Redshift Flag} (accepting flags beginning with `S', except those of unknown (`UN') or supernova (`SN') origin, and rejecting photometric measurements). If a published redshift uncertainty is not provided, we estimate one using the reported velocity uncertainty, or adopt a conservative fallback uncertainty of 1\%. 
    \item GLADE+ \citep{Glade+}, a comprehensive meta-catalogue of nearby galaxies constructed by combining multiple surveys (e.g., the Gravitational Wave Galaxy catalogue (GWGC; \citealt{White2011}), the 2MASS Extended Source catalogue (2MASS XSC; \citealt{2MASS, Bilicki2014}), and HyperLEDA (\citealt{Makarov2014})) to provide extensive sky coverage and high completeness for bright, low-redshift systems. We restrict the sample to entries with \texttt{dist\_flag}~=~3, corresponding to galaxies with spectroscopic redshifts.
\end{enumerate}

\begin{figure}
    \centering
    \includegraphics[width=\columnwidth]{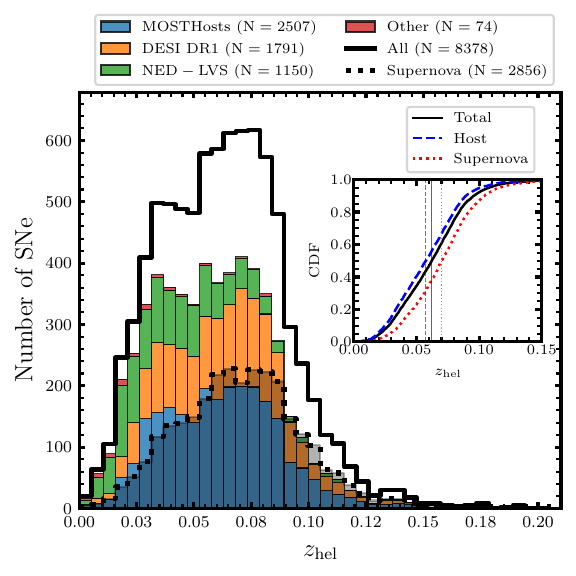}
    \caption{Distribution of best available heliocentric redshifts for the TITAN sample. The solid black curve shows the redshift distribution for all objects in the sample. The stacked, coloured histograms show distributions of the host galaxy redshifts, grouped by source (see Table \ref{tab:redshift_sources} for a complete description of the host-galaxy redshift sources; in the plot, `Other' comprises NED and GLADE+). The dashed black histogram shows the distribution of supernova redshifts, for which no host-galaxy redshift is available. The inset panel displays the cumulative distribution functions (CDFs) of the total sample, host-galaxy subsample, and supernova subsample; vertical lines mark the corresponding median redshifts of 0.062, 0.057, and 0.070, respectively.}
    \label{fig:redshift_dist}
\end{figure}

For each host galaxy, we perform a positional crossmatch between its central coordinates and entries in these catalogues within a 2$^{\prime\prime}$ radius. This process results in the successful acquisition of a spectroscopic host-galaxy redshift for 5522 SNe Ia (70.3\% of TITAN SNe Ia with an identified host, or 65.9\% of the full TITAN sample). Details of the individual redshift sources are presented in Table \ref{tab:redshift_sources}, and the TITAN redshift distribution is shown in Figure~\ref{fig:redshift_dist}.

\subsection{Redshifts from supernova spectral features} \label{sec:sn_z}

\begin{table}
\centering
\caption{Summary of best available redshift sources \rev{/ instruments} for all spectroscopically confirmed SNe Ia in the TITAN sample.}
\label{tab:redshift_sources}
\begin{tabular*}{\columnwidth}{@{\extracolsep{\fill}} l c c}
\hline
Redshift source & $N_\mathrm{SN}$ & Mean $\sigma_z$ \\
\hline
Host galaxy spectroscopy$^{a}$ & 5522 & $3.8\times10^{-5}$ \\
\quad MOSTHosts & 2507 & $1.1\times10^{-5}$ \\
\quad DESI DR1 & 1791 & $1.0\times10^{-5}$ \\
\quad NED-LVS & 1150 & $1.2\times10^{-4}$ \\
\quad NED & 58 & $3.1\times10^{-4}$ \\
\quad GLADE+ & 16 & $2.8\times10^{-4}$ \\
\texttt{SNID-SAGE} SN features & 2856 & $3.5\times10^{-3}$ \\
\quad \rev{SEDM} & \rev{1168} & \rev{$3.8\times10^{-3}$} \\
\quad \rev{EFOSC2-NTT} & \rev{660} & \rev{$3.3\times10^{-3}$} \\
\quad \rev{SNIFS} & \rev{314} & \rev{$3.5\times10^{-3}$} \\
\quad \rev{SPRAT} & \rev{148} & \rev{$3.2\times10^{-3}$} \\
\quad \rev{DBSP} & \rev{64} & \rev{$3.3\times10^{-3}$} \\
\quad \rev{KAST} & \rev{45} & \rev{$3.1\times10^{-3}$} \\
\quad \rev{ALFOSC} & \rev{44} & \rev{$3.0\times10^{-3}$} \\
\quad \rev{Other} & \rev{413} & \rev{$3.2\times10^{-3}$} \\
\hline
Total & 8378 & \\
\hline
\end{tabular*}
\vspace{2mm}
\begin{minipage}{\columnwidth}
\footnotesize
$^{a}$\,If a spectroscopic redshift exists for a galaxy hosting a SN with $d_\mathrm{DLR}>12$, it is included in the TITAN data release.
\end{minipage}
\end{table}

\begin{figure*}
    \centering
    \includegraphics[width=\textwidth]{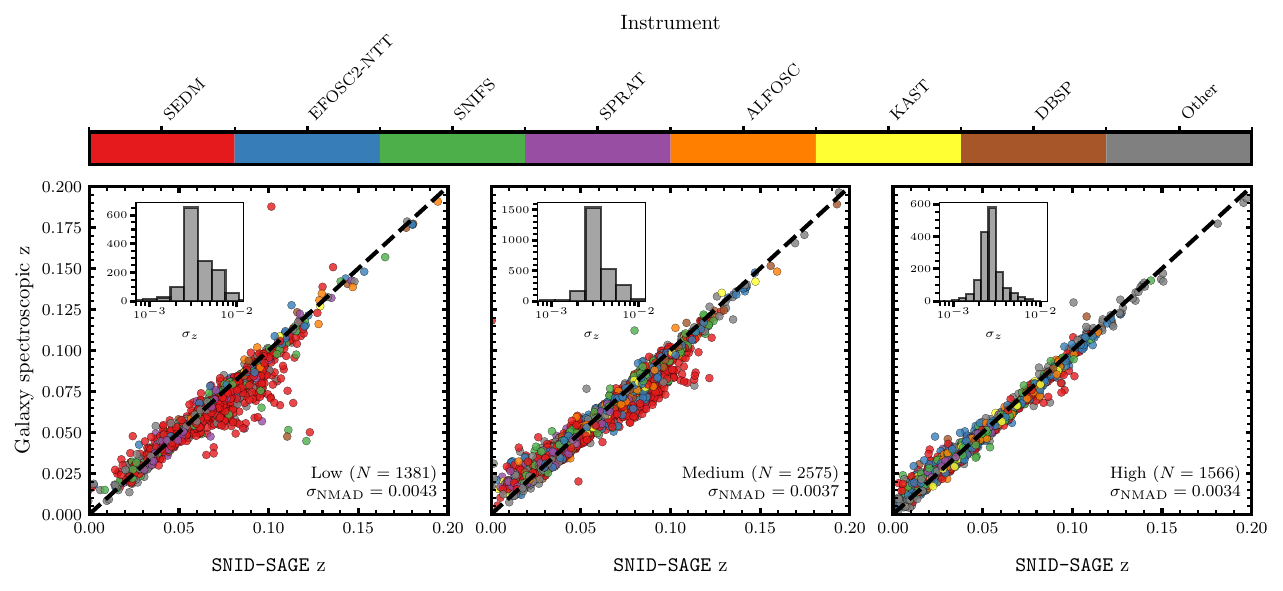}
    \caption{Comparison between redshifts derived from SN spectra using \texttt{SNID-SAGE} and the spectroscopic redshifts of the host galaxies. From left to right, each panel corresponds to a different SN–template match quality category (Low, Medium, High), with point colours indicating the instrument that produced the best-matching SN spectrum. Individual instruments are shown if they are responsible for $\geq100$ SN classifications, while the `Other' category encompasses all classifications by any other instruments. The black dashed line shows the one-to-one relation. Insets show the distributions of the estimated \texttt{SNID-SAGE} redshift uncertainties.}
    \label{fig:z_comp}
\end{figure*}

In instances where spectroscopic redshifts of the transient host galaxy are not available, the reported redshift is that obtained by template matching to the SN spectrum. All TITAN supernovae are re-classified using the \texttt{SNID-SAGE} spectral cross-correlation framework \citep{SNID-SAGE}, applied to publicly available spectra retrieved from \texttt{WISeREP} \citep{Wiserep} (see Murakami et al., \textit{in prep.}~for further details). \texttt{SNID-SAGE} follows a template-matching approach in which each observed spectrum is preprocessed, rebinned onto a logarithmic wavelength grid, and cross-correlated against a large library of rest-frame templates to identify the redshift that maximizes spectral feature alignment. 

For each spectrum, \texttt{SNID-SAGE} determines redshifts by aggregating the best-matching templates: candidate matches are ranked using a composite cross-correlation--based quality metric ($\mathrm{H}\sigma\mathrm{LAP-CCC}$) and split into redshift-space groups, with the highest-quality group adopted as the preferred solution. The quality of each group is quantified by a score $Q$, defined as the mean of its top five member $\mathrm{H}\sigma\mathrm{LAP-CCC}$ values (with a linear penalty applied when fewer than five templates contribute; \citealt{SNID-SAGE}), which is mapped onto discrete match-quality classes via fixed thresholds: \textit{Very Low} ($Q < 2.5$),\footnote[9]{\textit{Very Low} quality matches are excluded from the TITAN sample of definite SNe Ia (Murakami et al., \textit{in prep.})} \textit{Low} ($2.5 \leq Q < 5$), \textit{Medium} ($5 \leq Q < 8$), and \textit{High} ($Q \geq 8$). The final redshift is computed as a weighted mean of the contributing templates within the adopted group, with weights that scale exponentially with match quality, while the associated uncertainty is given by the weighted dispersion of these template redshifts, including a correction for the effective number of contributing templates.

To assess the reliability (and potential cosmological utility) of the redshifts derived from SN spectral features, Figure \ref{fig:z_comp} shows a direct comparison between spectroscopic galaxy redshifts and SN redshifts returned by \texttt{SNID-SAGE} as a function of match quality and instrument that produced the SN spectrum. As expected, the agreement between SN-$z$ and galaxy-$z$ improves markedly with increasing template-match quality: the high-quality sample lies tightly along the one-to-one relation, with little visible bias, whereas the medium-quality sample shows modest additional broadening and the low-quality sample exhibits substantially larger scatter and a much higher incidence of outliers. This trend is captured by the normalized median absolute deviation (nMAD) of the residuals, which decreases from $0.0043$ for the low-quality sample to $0.0037$ and $0.0034$ for the medium- and high-quality samples, respectively (when SEDM spectra are excluded, these scatter values reduce by 19\%, 11\% and 3\%, respectively). This behaviour is fully consistent with the \texttt{SNID-SAGE} validation presented by \cite{SNID-SAGE}. The inset histograms show that the \texttt{SNID-SAGE} redshift uncertainties, $\sigma_z$, concentrate towards smaller values as match quality improves. Though, as expected, few approach the precision of galaxy spectroscopic redshifts.

The instrument dependence visible in Figure \ref{fig:z_comp} is also informative. Concordantly with the findings of \cite{SNID-SAGE}, SEDM spectra  \citep{Blagorodnova2018} dominate the low- and medium-quality panels and are comparatively rarer among the high-quality matches, indicating that SEDM classifications more often produce less secure redshift solutions than spectra from instruments such as EFOSC2/NTT 
\citep{Smartt2015} or SNIFS \citep{Tucker2022}.  They also account for a significant fraction of the largest redshift deviations. In our sample, this effect appears particularly pronounced around $z \sim 0.06-0.12$, where a bulge of SEDM points departs from the one-to-one relation, suggesting that the resolving power of SEDM becomes insufficient to recover precise SN-based redshifts in this regime. Given that SEDM provides $\sim 42\%$ of the TITAN spectroscopic classifications (Murakami et al., \textit{in prep.}), this instrumental limitation is likely to be an important contributor to the overall SN-$z$ performance of the survey.

\rev{Of the 7568 hosts with successful global SED fits (Section~\ref{sec:fiducial_sed_fitting}), 2293 (30.2\%) rely on redshifts derived from SN spectral features, of which 968 are from SEDM spectra with no host-galaxy redshift available from any other source. Because the redshift is fixed in the SED fitting (Section~\ref{sec:fiducial_sed_fitting}), errors in these redshifts propagate into the derived host properties. We quantify this in Appendix~\ref{app:snz} by refitting \revtwo{the 2236} hosts that have both an SEDM-based and a host-galaxy spectroscopic redshift, fixing $z$ to the SN-based value. Stellar masses and SFRs shift in accordance with the change in luminosity distance, with a scatter of $\sim$0.06\,dex for the typical SEDM redshift error \revtwo{and only 81 hosts (3.6\%) moving across the conventional mass-step boundary at $\log_{10}(M_*/M_\odot) = 10$}, while distance-independent properties (e.g., $A_V$, mass-weighted age) are essentially unaffected.}

These trends have direct implications for cosmological analyses, and future work will investigate how including SNe without host-$z$ measurements but with high-quality \texttt{SNID-SAGE} SN-$z$ values affects the scatter in the TITAN Hubble diagram.

\subsection{Peculiar velocity corrections}

Observed redshifts of nearby galaxies are influenced by several velocity components. These include the motion of the observer relative to the heliocentric frame, the motion of the Solar System with respect to the cosmic microwave background (CMB) rest frame (i.e., the CMB dipole), and the peculiar velocities (PVs) of the galaxies themselves arising from local gravitational flows. For precision cosmological analyses, these contributions must be accounted for by transforming redshifts into the CMB rest frame and correcting for host-galaxy PVs, thereby reducing biases and excess scatter in the low-redshift Hubble diagram.

Observed redshifts, $z_\mathrm{obs}$, are first transformed to the heliocentric frame, yielding $z_\mathrm{hel}$, by correcting for the motion of the observer due to both the Earth's rotation and its orbit around the Sun. This correction depends on the observatory location, the time of observation, and the sky position of the target. In practice, however, the galaxy redshift catalogues used in this work report values that have already been corrected to the heliocentric frame. We therefore take all tabulated redshifts used in this work to correspond to $z_\mathrm{hel}$.

To transform heliocentric redshifts into the CMB rest frame, we correct for the peculiar motion of the Solar System inferred from the CMB dipole. We adopt the \cite{Planck2020} measurement of a Solar System velocity of $v_\mathrm{Sun}^\mathrm{max} = (369.82 \pm 0.11)\,\mathrm{km\,s^{-1}}$ towards Galactic coordinates $(l, b) = (264.021^\circ \pm 0.011^\circ,\ 48.253^\circ \pm 0.005^\circ)$. The component of this velocity along the line of sight to an object is
\begin{equation}
    v_\mathrm{Sun} = v_\mathrm{Sun}^\mathrm{max} \cos \alpha,
\end{equation}
where $\alpha$ is the angular separation between the object and the CMB dipole apex. The corresponding relativistic redshift induced by this motion is
\begin{equation}
    1 + z_\mathrm{Sun} = \sqrt{\frac{1 - v_\mathrm{Sun}/c}{1 + v_\mathrm{Sun}/c}},
\end{equation}
and the redshift in the CMB rest frame is therefore given by
\begin{equation}
    1 + z_\mathrm{CMB} = \frac{1 + z_\mathrm{hel}}{1 + z_\mathrm{Sun}}.
\end{equation}

To obtain the final, cosmologically useful redshift, $z_\mathrm{HD}$, we must additionally correct for the PVs of the SN host galaxies. In the following sections, we describe two independent methods for estimating these PV corrections, parametrized as an effective redshift term, $z_\mathrm{PV}$. The resulting Hubble diagram redshift is then given by
\begin{equation}
    1 + z_\mathrm{HD} = \frac{1 + z_\mathrm{CMB}}{1 + z_\mathrm{PV}}.
\end{equation}
For completeness, the TITAN data release provides all relevant redshift quantities, including $z_\mathrm{hel}$, $z_\mathrm{CMB}$, and $z_\mathrm{HD}$.

\subsubsection{\texttt{\upshape Manticore-Local} peculiar velocities}

Our baseline peculiar velocity estimates are derived from the \texttt{Manticore-Local} project \citep{Manticore} -- a high-resolution digital twin of the local Universe that self-consistently models structure formation and galaxy motions. Following \cite{stiskalekVelocityFieldOlympics2025}, the posterior of the cosmological redshift $z_\text{HD}$ of a galaxy at a redshift-space position $z_\mathrm{CMB}$ is obtained by integrating the following probability
\rev{
\begin{multline}
      \mathcal{P}(z_\text{HD}|z_\mathrm{CMB}) \propto \idotsint dv\,d\rho dV_\mathrm{ext}\,d\sigma_v\,d\alpha_\text{low}\,d\alpha_\text{high}\,d\rho_t \\
      \mathcal{P}(z_\mathrm{CMB}|v,z_\text{HD}, V_\mathrm{ext}, \sigma_v)\mathcal{\pi}(z_\text{HD}|\rho, \rho_t, \alpha_{\rm low}, \alpha_{\rm high})
     \\\times\mathcal{P}(v,\rho, V_\mathrm{ext},\sigma_v, \rho_t, \alpha_{\rm low}, \alpha_{\rm high}),
      \label{eq:vpost}
\end{multline}}
where $V_{\rm ext}$ is the velocity of the bulk-flow out of the reconstructed volume, $\sigma_v$ is the non-linear velocity noise, and $\alpha_{\rm low/high}$, $\rho_t$ are the parameters of the inhomogeneous Malmquist bias double power-law as presented in \cite{stiskalekVelocityFieldOlympics2025}. The cosmological redshift distribution $\mathcal{P}(z_\mathrm{CMB}|v,z_\text{HD}, V_\mathrm{ext}, \sigma_v)$ is given by
\begin{align}
        \mathcal{P}(z_\mathrm{CMB}|v,z_\text{HD}, V_\mathrm{ext}, \sigma_v) \sim \mathcal{N}(\mu, \sigma)\\ \text{ with } \left\{
   \begin{matrix}
        \mu &=& \left(1+z_\mathrm{HD}\right)\left(1 + \left(v + V_\mathrm{ext}\right) / c\right) - 1\\
        \sigma &=& \sqrt{\sigma_v^2 + \sigma_{z_\mathrm{CMB}}^2}
    \end{matrix}
    \right.
\end{align}
and $\mathcal{\pi}(z_\text{HD}|\rho)$ accounts for both homogeneous and inhomogeneous Malmquist bias using a double-power law, given by \citep{stiskalekVelocityFieldOlympics2025}
\begin{multline}\label{eq:malmprior}
    \mathcal{\pi}(z_\text{HD}|\rho, \rho_t, \alpha_{\rm low}, \alpha_{\rm high}) \\
    =  \frac{1}{\mathcal{Z_\pi}} r^2\left(\frac{\rho}{\rho_t}\right)^{\alpha_\text{low}} \left(1 + \frac{\rho}{\rho_t}\right)^{\alpha_\text{high} - \alpha_\text{low}}\frac{\partial r}{\partial z_\text{HD}},
\end{multline}
where $\mathcal{Z_\pi}$ is the normalization factor of Eq.~\ref{eq:malmprior}.

The integral over the probability distribution $\mathcal{P}(v,\rho, V_\mathrm{ext},\sigma_v, \rho_t, \alpha_\mathrm{low}, \alpha_\mathrm{high})$ in Eq~\ref{eq:vpost} is obtained by averaging over $N=80$ samples of the posterior of \texttt{Manticore-Local} and $M=5000$ samples of the posterior of a calibration of \texttt{Manticore-Local} using 2MTF data \citep{Hong2019} within the CANDEL\footnote[10]{\url{https://github.com/Richard-Sti/CANDEL}.} framework \citep{stiskalekVelocityFieldOlympics2025}.

To estimate the velocities, we compute numerically, for each pair $(m,n)$ of the \texttt{Manticore-Local} and calibration posteriors, the normalization $\mathcal{Z}$ of the second term of Eq.~\ref{eq:vpost}, and the first and second moment of the estimated velocity
\begin{align}
   I_{m,n}^{(1)} = \frac{1}{\mathcal{Z}}  \int d z_\text{HD}  v_p \mathcal{\pi}(z_\text{HD}|\rho)\mathcal{P}(z_\mathrm{CMB}|v,z_\text{HD}, V_\mathrm{ext}, \sigma_v),
\end{align}
and,
\begin{align}
   I_{m,n}^{(2)}  
   = \frac{1}{\mathcal{Z}}  \int d z_\text{HD}  v_p^2 \mathcal{\pi}(z_\text{HD}|\rho)\mathcal{P}(z_\mathrm{CMB}|v,z_\text{HD}, V_\mathrm{ext}, \sigma_v),
\end{align}
where the total velocity on the line of sight is defined as
\begin{equation}
    v_p = c\left(\frac{1+z_\mathrm{CMB}}{1+z_\mathrm{HD}}-1\right).
\end{equation}
The velocity is then obtained as
\begin{equation}
    \hat{v}_p = \frac{1}{NM} \sum_{i, j = 1,1}^{m, n} I_{i,j}^{(1)}.
\end{equation}
The error on the velocity is computed following the law of total variance as
\begin{align}
   \sigma_{v_p}^2 = \frac{1}{NM}\sum_{m,n} \sigma^2_{v_p, m,n} + \frac{1}{NM}\sum_{m,n} \left(I_{m,n}^{(1)} - \hat{v}_p\right)^2,
\end{align}
where $\sigma^2_{v_p, m,n}$ is the inner realisation variance, given by
\begin{equation}
    \sigma^2_{v_p, m,n} = I^{(2)}_{m,n} - \left(I_{m,n}^{(1)}\right)^2.
\end{equation}

\subsubsection{2M++ peculiar velocities}

We additionally derive peculiar velocities following the process outlined in \citet{Peterson2022} where bulk-flow corrections are calculated using a velocity-field reconstruction from 2M++ data \citep{Lavaux2011}.
In constructing the peculiar velocity field, we follow both \citet{Carrick2015} and \citet{Carr2022} by converting the velocity field into real space coordinates and then into redshift space.
We also include an external bulk-flow correction from SDSS \citep{Albareti2017} as detailed in \citet{Said2020} (this is the `2M++/SDSS' bulk-flow correction described in section 4.1 of \citealt{Peterson2022}).
For this analysis, we implement this process by using the \texttt{pvhub} tool.\footnote[11]{\url{https://github.com/KSaid-1/pvhub}.} These alternative PV estimates are included as an independent systematic check for future TITAN cosmological analyses, enabling direct comparisons between different velocity-field reconstructions and assessments of their impact on low-redshift Hubble residuals and cosmological parameter inference. The same 2M++ reconstruction \citep{Carrick2015} underpins the growth-of-structure ($f\sigma_8$) forecast for TITAN presented by \citet{Dixon2026}, who validate the reconstruction-and-scaling method and quantify its first full systematic uncertainty budget against the Uchuu $N$-body simulations \citep{Ishiyama2021}.

\section{Photometry}\label{sec:photometry}

\begin{table}
\centering
\caption{Photometric coverage of the 7568 TITAN host galaxies with successful global SED fits. For each band we give the number and percentage of hosts with a usable measurement entering the SED fit (`Measured'), irrespective of signal-to-noise and including negative fluxes consistent with zero, and the number and percentage detected at S/N\,$>3$ (`Detected'). Failed measurements, as described in Section~\ref{sec:phot_uncertainties}, are excluded. For the wavelength regimes, `Detected' requires a detection in at least one band of each regime (or in every band, where stated). The final row gives the median number of bands per galaxy, with the 16th--84th percentile range in parentheses.}
\label{tab:coverage}
\setlength{\tabcolsep}{4pt}
\begin{tabular}{llrrrr}
\hline
 & & \multicolumn{2}{c}{Measured} & \multicolumn{2}{c}{Detected} \\
\cline{3-4}\cline{5-6}
Survey & Band & $N$ & \% & $N$ & \% \\
\hline
GALEX & FUV & 6027 & 79.6 & 4429 & 58.5 \\
 & NUV & 6398 & 84.5 & 6255 & 82.7 \\
\noalign{\smallskip}
SDSS & $u$ & 4462 & 59.0 & 3129 & 41.3 \\
 & $g$ & 4453 & 58.8 & 4013 & 53.0 \\
 & $r$ & 4460 & 58.9 & 4189 & 55.4 \\
 & $i$ & 4453 & 58.8 & 4094 & 54.1 \\
 & $z$ & 4447 & 58.8 & 4095 & 54.1 \\
\noalign{\smallskip}
Pan-STARRS & $g$ & 7158 & 94.6 & 7134 & 94.3 \\
 & $r$ & 7161 & 94.6 & 7150 & 94.5 \\
 & $i$ & 7158 & 94.6 & 7153 & 94.5 \\
 & $z$ & 7158 & 94.6 & 7127 & 94.2 \\
 & $y$ & 7147 & 94.4 & 6975 & 92.2 \\
\noalign{\smallskip}
DES & $g$ & 779 & 10.3 & 777 & 10.3 \\
 & $r$ & 782 & 10.3 & 781 & 10.3 \\
 & $i$ & 781 & 10.3 & 779 & 10.3 \\
 & $z$ & 784 & 10.4 & 782 & 10.3 \\
 & $Y$ & 773 & 10.2 & 764 & 10.1 \\
\noalign{\smallskip}
Legacy Survey (south) & $g$ & 3893 & 51.4 & 3844 & 50.8 \\
 & $r$ & 3644 & 48.2 & 3618 & 47.8 \\
 & $z$ & 3819 & 50.5 & 3776 & 49.9 \\
\noalign{\smallskip}
Legacy Survey (north) & $g$ (BASS) & 1646 & 21.7 & 1633 & 21.6 \\
 & $r$ (BASS) & 1649 & 21.8 & 1643 & 21.7 \\
 & $z$ (MzLS) & 1646 & 21.7 & 1625 & 21.5 \\
\noalign{\smallskip}
2MASS & $J$ & 7398 & 97.8 & 4638 & 61.3 \\
 & $H$ & 7400 & 97.8 & 3789 & 50.1 \\
 & $K_s$ & 7388 & 97.6 & 3440 & 45.5 \\
\noalign{\smallskip}
WISE & W1 & 7431 & 98.2 & 6664 & 88.1 \\
 & W2 & 7429 & 98.2 & 6310 & 83.4 \\
 & W3 & 7432 & 98.2 & 4035 & 53.3 \\
 & W4 & 7392 & 97.7 & 1218 & 16.1 \\
\hline
\multicolumn{6}{l}{\textit{Wavelength regime}} \\
\multicolumn{2}{l}{Any optical} & 7568 & 100.0 & 7565 & 99.9 \\
\multicolumn{2}{l}{Any UV (FUV or NUV)} & 6479 & 85.6 & 6318 & 83.5 \\
\multicolumn{2}{l}{Any NIR ($JHK_s$)} & 7428 & 98.2 & 4736 & 62.6 \\
\multicolumn{2}{l}{Any MIR (W1--W4)} & 7446 & 98.4 & 6697 & 88.5 \\
\multicolumn{2}{l}{Optical, FUV, NUV, and W3 or W4$^{a}$} & 5842 & 77.2 & 2847 & 37.6 \\
\multicolumn{2}{l}{Full coverage$^{b}$} & 5621 & 74.3 & 629 & 8.3 \\
\hline
\multicolumn{2}{l}{Bands per galaxy} & \multicolumn{2}{c}{20 (14--22)} & \multicolumn{2}{c}{16 (11--20)} \\
\hline
\end{tabular}
\medskip
\begin{minipage}{\columnwidth}
\footnotesize
$^{a}$ The sample for which SFRs, $A_V$ and $t_{\rm MWA}$ are reported, subject to the removal of AGN contamination (Section~\ref{sec:excluding_phot}).\\
$^{b}$ At least one set of optical bands and all GALEX, 2MASS and WISE bands; `Detected' requires a detection in every GALEX, 2MASS and WISE band.
\end{minipage}
\end{table}
We here outline the procedure used to obtain the multiband aperture photometry of each host galaxy in the TITAN sample, which is then used in the SED fitting procedure outlined in the next section. We obtain both global photometry, wherein the aperture encapsulates the entirety of the host galaxy, and local photometry, where fluxes are measured within an aperture of $1.5\ \mathrm{kpc}$ centred on the SN location (see Section \ref{sec:local_photometry} for justification of this aperture size).

Multiband photometric measurements for all host galaxies in the TITAN sample are obtained using the \texttt{HostPhot} package \citep{MullerBravo2022}. Images from a subset or all of the following surveys and bandpasses were used, depending on sky coverage for a given region: Pan-STARRS DR1 (\textit{grizy}; \citealt{PanSTARRS}), the Dark Energy Survey (DES) DR2 (\textit{grizY}; \citealt{DES_DR2}), SDSS DR17 (\textit{ugriz}; \citealt{SDSS_DR17}), the DESI Legacy Imaging Surveys DR10 (BASS \textit{gr} + MZLS \textit{z}; \citealt{LegacySurvey}), the Galaxy Evolution Explorer (GALEX) (\textit{FUV, NUV}; \citealt{GALEX}),\footnote[12]{Each of the Deep (DIS), Medium (MIS), All-Sky Imaging (AIS), Nearby Galaxy (NGS) and Guest Investigator (GII) surveys are queried; the image with the longest exposure time is selected.} 2MASS (\textit{JHK$_s$}; \citealt{2MASS}), and unWISE (\textit{W1–W4}; \citealt{unWISE}).\footnote[13]{We use the AllWISE version of the unWISE images, which includes all four WISE bands; the NeoWISE reactivation data provide only \textit{W1} and \textit{W2}.} \rev{Images from all surveys other than 2MASS and unWISE are already background-subtracted; for these two, \texttt{HostPhot} performs the background subtraction itself.}

A total of 128 galaxies (1.6\% of those with a confident host association) lack coverage in any of the available optical imaging surveys and are therefore excluded from the photometric analysis, as we require optical data to define source masks and apertures, and to form the basis of the SED fitting outlined in Section \ref{sec:sed_fitting}. The remaining systems proceed through the full photometric pipeline described in the following subsections.

\rev{The photometric coverage of the sample is summarised in Table~\ref{tab:coverage}. We consider a band to be covered when a measurement is available within the global aperture, irrespective of its signal-to-noise; non-detections, including negative fluxes consistent with zero, are retained in the SED fitting (Section~\ref{sec:phot_uncertainties}). The median host has usable measurements in 20 bands (16th--84th percentile range 14--22), of which 16 (11--20) are detected at S/N\,$>3$. Coverage is essentially complete in Pan-STARRS (95\%), 2MASS and WISE (98\%). In the UV, $\sim14\%$ of hosts lack any GALEX photometry, of which 77\% lie outside the footprint of the GALEX All-sky Imaging Survey (predominantly at low Galactic latitude, which the survey avoided) and most of the remainder lie very close to its edges; a further 5\% have NUV photometry only, the majority of which lie outside the FUV footprint, reflecting the loss of the FUV detector in 2009. The DES and Legacy Survey (DR10, south) images are taken with the same camera and filters; where both are available for a host, we adopt the deeper DES photometry.}

\subsection{Global photometry} \label{sec:global_photometry}

Accurate global photometry relies on isolating the flux of the host galaxy from contaminating sources within the field of view. In crowded or highly resolved low-redshift systems, foreground stars, background galaxies, and image artefacts can bias flux measurements if not properly accounted for. We therefore first construct source masks to exclude non-host emission, forming the basis for subsequent aperture definition and photometric measurements.

\subsubsection{Masking} \label{sec:masking}

For each host, co-added images are first generated from the available optical filters (\textit{riz} for Pan-STARRS, DES and SDSS, and \textit{rz} for Legacy Survey), and these are used to define source masks that exclude objects other than the target host galaxy. All masks are visually inspected to ensure that host galaxy emission is preserved while contaminants are effectively removed. Where necessary, masking parameters are iteratively adjusted on a per-object basis. The final masks are saved and propagated to the individual single-filter images across all surveys, ensuring consistent exclusion of contaminating sources in the subsequent photometric measurements.

A total of 61 galaxies (0.8\%) fail this step. This can occur for a variety of reasons. In some cases, the host resides in an interacting or merging system, where the supernova can be confidently associated with one component, but the galaxy itself cannot be cleanly separated from its companion for reliable photometry. In other instances, a substantial fraction of the host galaxy is obscured or masked by a bright foreground star, diffraction spikes, or a nearby extended source, leading to significant loss of flux. Additional failures arise in crowded fields where aggressive masking is required to remove contaminants, but at the expense of removing genuine host emission, or in systems with complex or irregular morphologies where automated source detection struggles to distinguish between host structure and neighbouring objects. In a small number of cases, image artefacts or defects (e.g., scattered light, saturation, or poor background subtraction) prevent the construction of a reliable mask. These systems are excluded from further photometric analysis by visual inspection, as their photometry would be systematically biased and not suitable for robust SED fitting.

\subsubsection{\rev{Apertures and Flux Measurements}} \label{sec:apertures}

\begin{figure}
    \centering
    \includegraphics[width=\columnwidth]{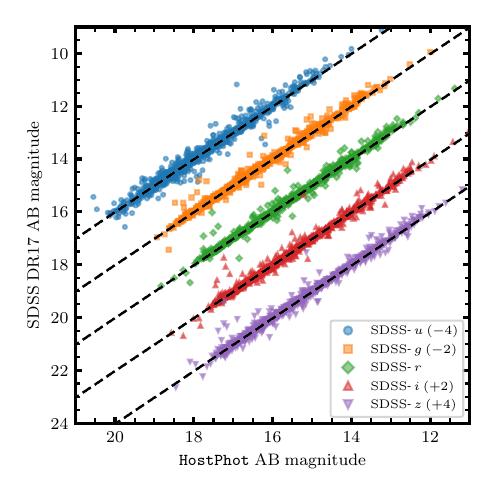}
    \caption{Comparison between SDSS DR17 catalogue fluxes and fluxes derived from \texttt{HostPhot} for the same galaxies in the SDSS \textit{ugriz} bands. Points are coloured by band, and are offset to improve visibility. Dashed black lines show 1-to-1 relations, demonstrating strong consistency between the measured fluxes.}
    \label{fig:sdss_flux_comparison}
\end{figure}

\begin{figure}
    \centering
    \includegraphics[width=\columnwidth]{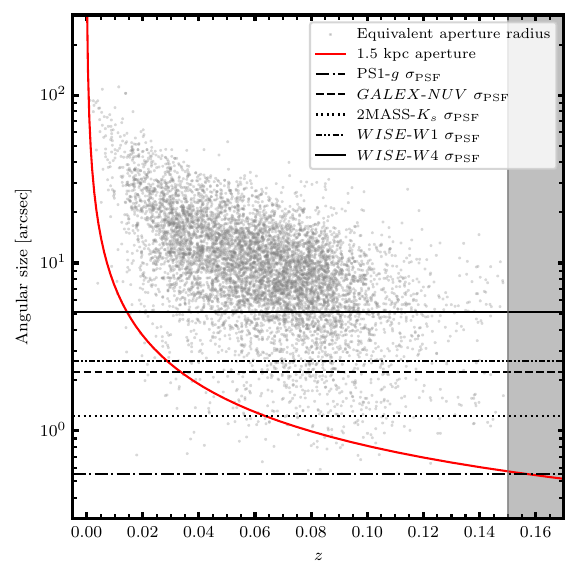}
    \caption{The apparent angular size of representative \rev{global (optical)} apertures as a function of redshift. The red curve shows the angular radius corresponding to a fixed physical scale of 1.5 kpc. The vertical grey region marks the redshift range over which this aperture becomes smaller than the Pan-STARRS $\sigma_\mathrm{PSF}$, and local photometry is therefore not performed. Horizontal lines indicate the worst-case PSF size \rev{($\sigma_\mathrm{PSF}$)} in key bands from the surveys used in this work. Grey points show the radii of circular apertures with areas equivalent to the elliptical apertures adopted for global photometry, illustrating the typical angular scales probed across the sample.}
    \label{fig:ap_size_vs_z}
\end{figure} 

\begin{figure*}
    \centering
    \includegraphics[width=\textwidth]{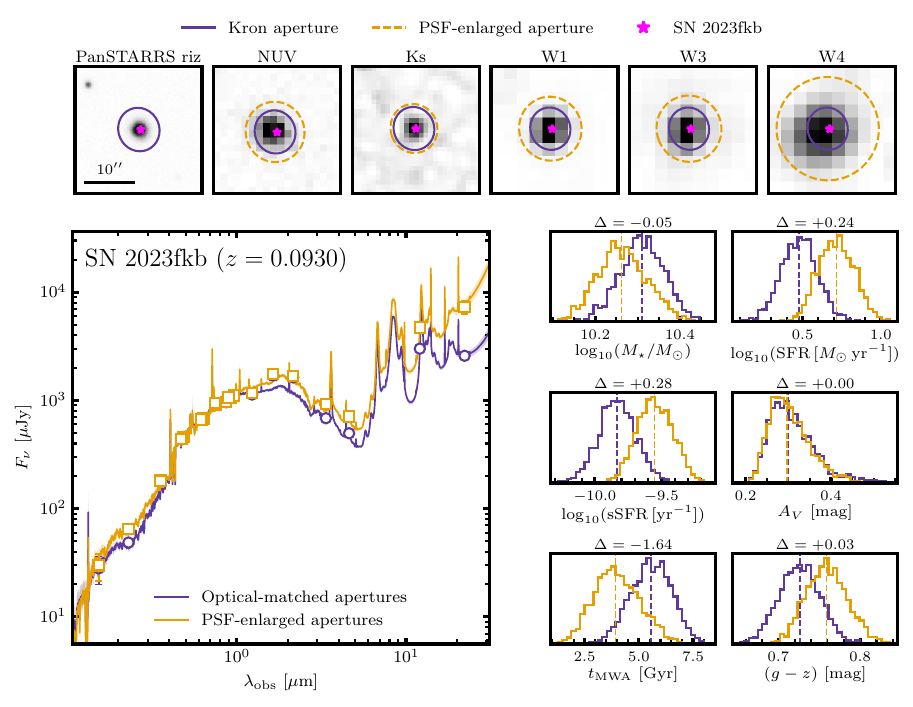}
    \caption{\rev{Effect of the PSF-enlarged apertures for the host of SN\,2023fkb ($z = 0.093$), a compact host with an equivalent aperture radius of 4.5$^{\prime\prime}$. \textit{Top:} cutouts of the host in the Pan-STARRS $riz$ co-add, GALEX NUV, 2MASS $K_s$ and WISE W1, W3 and W4, all on the same angular scale, showing the Kron aperture defined on the optical co-add (\revtwo{solid purple}) and the aperture enlarged in quadrature with $2\,\sigma_{\rm PSF}$ in each band (\revtwo{dashed orange}); the magenta star marks the SN position. \textit{Bottom left:} posterior median spectra (lines) and 16th--84th percentile ranges (shaded) from fits using apertures matched to the optical (\revtwo{purple}) and the PSF-enlarged apertures (\revtwo{orange}), with the corresponding photometry (\revtwo{open circles for optical, open squares for PSF-enlarged}). \textit{Bottom right:} posterior distributions of the stellar mass, SFR, sSFR, $A_V$, mass-weighted age and rest-frame $g-z$ colour for the two fits, with the posterior medians shown as dashed lines and the difference between them, $\Delta$ (enlarged minus optical-matched), given above each panel.}}
    \label{fig:enlarged_example}
\end{figure*}

Using the masked optical co-added images, we determine an adaptive elliptical aperture for each host galaxy following the Kron radius formalism \citep{Kron1980}, as implemented in \texttt{HostPhot}. Source detection is performed on the co-add (using the \textit{riz} or \textit{rz} stack), and the initial aperture parameters (centroid, semi-major axis, axis ratio, and position angle) are derived from the light distribution of the detected source. The aperture size is then iteratively optimized by expanding the Kron radius until the fractional change in enclosed flux falls below a convergence threshold (fiducially $\epsilon = 5 \times 10^{-4}$), ensuring that the majority of the galaxy light is captured while minimizing contamination from the background.

Each aperture is subsequently visually inspected \rev{on the optical co-adds} to confirm that the host galaxy is fully enclosed and that the derived shape parameters are appropriate. In cases where the automated procedure fails (e.g., for irregular or low-surface-brightness systems), the aperture scale and/or convergence threshold are manually adjusted and recomputed. \rev{A total of 99 (1.3\%) galaxies fail this step, in the majority of cases because the host is too faint to be robustly detected by \texttt{sep} in the optical co-adds, preventing a stable determination of the Kron radius. These objects -- together with the 128 galaxies lacking optical coverage and the 61 that fail the masking step -- do not have well-defined light profiles or apertures, and no meaningful $d_{\rm DLR}$ can be assigned to them. In total, 7568 of the 7856 (96.3\%) confidently identified host galaxies have well-measured apertures and are carried forward to the SED fitting stage. The final aperture parameters are then fixed, and the same elliptical aperture, with identical centre, orientation and axis ratio, is propagated to every available band.}

\rev{To validate the optical photometry, Figure~\ref{fig:sdss_flux_comparison} compares our \texttt{HostPhot} fluxes with those of the SDSS DR17 catalogue \citep{SDSS_DR17} for the TITAN hosts in common. Although the SDSS fluxes are measured within elliptical Petrosian apertures, the two agree closely in all $ugriz$ bands, with nMAD $\leq0.05$\,mag, demonstrating that our optical photometry is accurate and that the choice of aperture definition introduces no significant systematic offset.}

\rev{Because the fraction of a source's flux enclosed by a fixed aperture depends on the PSF, applying an identical aperture to bands of very different resolution would systematically underestimate the flux in the lowest-resolution bands for compact hosts. The PSF FWHM ranges from $\lesssim$1.3$^{\prime\prime}$ in the optical to $\sim$4--5$^{\prime\prime}$ in GALEX \citep{GALEX_calibration}, $\sim$3$^{\prime\prime}$ in 2MASS \citep{2MASS}, and 6--12$^{\prime\prime}$ in WISE \citep{Wright2010}, comparable to or larger than the optically-derived apertures of some TITAN hosts. In Figure~\ref{fig:ap_size_vs_z}, we show that 18.3\% of hosts have optical apertures, expressed as the radius of a circle of equal area, smaller than $\sigma_{\rm PSF} = {\rm FWHM}/2\sqrt{2\ln2}$ in WISE W4, and 5.8\% in W1. We therefore enlarge the aperture \revtwo{of every host} in the GALEX, 2MASS and WISE bands in quadrature with the PSF of each band, replacing the semi-axes $a$ and $b$ of the Kron ellipse with $\sqrt{a^2 + (k\,\sigma_{\rm PSF})^2}$ and $\sqrt{b^2 + (k\,\sigma_{\rm PSF})^2}$. As PSF-convolved profiles broaden in quadrature, this approximately matches the fraction of each host's light enclosed across bands, rather than the angular aperture, following the principle of PSF-convolved apertures \citep[e.g.,][]{Wright2016}. Extended hosts ($a,b \gg k\,\sigma_{\rm PSF}$) are essentially unaffected, while for unresolved hosts the aperture tends to a radius of $k\,\sigma_{\rm PSF}$. We adopt $k = 2$, for which this limiting aperture encloses 86\% of the flux of a point source, comparable to the fraction of light captured by Kron apertures in the optical. Larger values recover only fractionally more flux, at a substantial cost in signal-to-noise in the background-dominated W3 and W4 bands, whose noise is correlated across the aperture and therefore grows faster than the square root of the aperture area, and with an increased risk of contamination by neighbouring sources.}

\rev{Figure~\ref{fig:enlarged_example} illustrates this for the host of SN\,2023fkb, whose aperture (equivalent radius 4.5$^{\prime\prime}$) is comparable to $\sigma_{\rm PSF}$ in W4. The enlarged apertures increase the W3 and W4 fluxes by factors of 1.6 and 2.8, respectively, and the GALEX and NIR fluxes by smaller amounts, while the optical photometry is unchanged. The fit with the enlarged apertures yields an SFR higher by 0.24\,dex and a mass-weighted age younger by 1.6\,Gyr, with the stellar mass changing by only $-0.05$\,dex. $A_V$ is essentially unchanged, as the UV and MIR fluxes increase together, preserving the ratio of absorbed to escaping starlight that constrains the attenuation within the energy-balance framework; the additional emission is instead attributed to a younger, more strongly star-forming population.}

Finally, each SED is visually inspected prior to fitting to identify any spurious photometric measurements not flagged during \rev{previous} steps. Such cases typically arise either from image artefacts that spuriously inflate the measured fluxes, or from edge effects, where a galaxy near the boundary of a survey footprint is only partially covered and its flux is therefore underestimated. Problematic measurements are \rev{then excluded} before proceeding with the SED fitting. \rev{This process results in the removal of 749 photometric measurements across 328 galaxies (0.4\% of all photometry).} \rev{All remaining measurements} are corrected for foreground Galactic extinction using the dust maps of \citet{SchlaflyFinkbeiner2011} in combination with the extinction law of \citet{Fitzpatrick1999} with $R_V=3.1$. We additionally adopt a 5\% error floor on all photometric measurements to account for systematic uncertainties, including residual calibration offsets, background subtraction errors, and \rev{residual differences in the fraction of flux enclosed across bands}.

\subsubsection{\rev{Non-detections and photometric uncertainties}}
\label{sec:phot_uncertainties}
 
\rev{Here we describe additional treatment of the photometric uncertainties beyond that provided by \texttt{HostPhot}, together with our treatment of non-detections. Non-detections are retained in the SED fitting as flux measurements with their associated uncertainties, including negative fluxes consistent with zero.}
 
\rev{For 2MASS and WISE, for which \texttt{HostPhot}\footnote[14]{\revtwo{This work makes use of \texttt{HostPhot} v3.0.0rc0.}} performs the background subtraction, the returned uncertainties do not appear to fully capture the uncertainty in the local background level. This manifests as an excess of significantly negative fluxes among undetected hosts: before correction, 8--27\% of negative 2MASS and W3/W4 measurements lie below $-3\sigma$, compared with 0.3\% expected for Gaussian noise. These measurements arise predominantly among faint hosts with apertures of $r_\mathrm{ap}\sim4-8^{\prime\prime}$. We therefore add in quadrature an additional term, $\sigma_{\rm bkg} = c_{\rm band}\,A_{\rm ap}$, where $A_{\rm ap}$ is the area of the PSF-enlarged aperture and $c_{\rm band}$ is fitted for each band by maximum likelihood, requiring that the negative-flux measurements follow a half-normal distribution of unit width. The fitted coefficients are 1.14, 1.94 and 1.96\,$\mu$Jy\,arcsec$^{-2}$ for 2MASS $J$, $H$ and $K_s$, and 0.07, 0.06, 0.28 and 0.52\,$\mu$Jy\,arcsec$^{-2}$ for WISE W1--W4, corresponding to median error inflations of 2.9--3.6 for 2MASS and $\leq$1.2 for WISE. After this correction, 0.5--1.9\% of negative-flux measurements lie below $-3\sigma$, with no residual dependence on aperture size. Measurements that remain below $-3\sigma$ are treated as failed measurements, and are excluded from the SED fitting; this removes 98 of the $\sim$52\,000 2MASS and WISE measurements for the fitted hosts, while all other measurements, including negative fluxes consistent with zero, are retained.}
 
\rev{We validate this correction against the 2MASS XSC \citep{2MASS, Bilicki2014} for the 3894 hosts matched to catalogued galaxies within $5^{\prime\prime}$. Modelling the scatter between our fluxes and the XSC total fluxes in bins of aperture radius, we find that bright, extended hosts require only $\sim$60\% of the amplitude of the added term beyond $r_{\rm ap} \sim 10^{\prime\prime}$. Because the XSC preferentially contains hosts in uncontaminated fields, this comparison does not probe the fainter hosts and more problematic fields that drive the excess of negative fluxes, for which the full amplitude is required. We therefore retain the calibrated coefficients, noting that 2MASS measurements of bright hosts are conservatively down-weighted in the SED fits; given the limited leverage of NIR data on the derived properties (Section~\ref{sec:excluding_phot}), this has negligible impact. The median ratio of our fluxes to the XSC total fluxes is 0.96--1.01 for $r_{\rm ap} \gtrsim 5^{\prime\prime}$, decreasing to $\sim$0.82 for the most compact hosts, consistent with the $\sim$86\% of the flux of a point source enclosed by the PSF-enlarged aperture (Section~\ref{sec:apertures}).}

\subsection{Local photometry} \label{sec:local_photometry}

While global properties of SN~Ia host galaxies are demonstrably useful proxies for correcting their distance moduli in cosmological analyses, these quantities represent averages over entire galaxies and may therefore obscure the specific physical conditions most relevant to the progenitor system. A more direct probe of the supernova birth environment can be obtained by measuring the properties of the region immediately surrounding the SN location (typically on kiloparsec scales), where the stellar population age, star-formation activity, metallicity, and dust content are more closely connected to progenitor evolution pathways. 

Ideally, one would seek to characterize the full range of physical properties locally -- beyond simple proxies such as stellar mass or integrated colour -- in analogy with global host measurements. 
In practice, such measurements require spatially resolved spectroscopy \rev{\citep[e.g.,][]{Rigault2013, Galbany2014, Galbany2018, Rigault2020}}, which is observationally demanding and not widely available for large samples. In this work we instead employ aperture photometry, following a similar methodology to \citet{Kelsey2021}, to estimate local properties, acknowledging the limitations imposed by angular resolution and PSF effects. We here define the local aperture size used in our analysis, and assess its suitability across the redshift range of the sample and the range of imaging data employed. Local aperture photometry is attempted for the 7568 galaxies that possess reliable global apertures\rev{; a more in depth study of the environments of orphan SNe and those with ambiguous or ultra-faint hosts is reserved for future work}.

\subsubsection{Aperture size}
\label{sec:local_aperture}

\rev{Local apertures must be large enough to resolve structure in every band used, and small enough to probe genuinely local environments. At the upper redshift limit of our sample, $z = 0.15$, the PSF widths of the GALEX, 2MASS and WISE bands (Section~\ref{sec:apertures}) correspond to minimum useful aperture radii of $\sim$3--13\,kpc, far beyond the kiloparsec scales over which stellar population properties vary within galaxies (e.g., \citealt{Sanchez2012, Sanchez2014}; \citealt{GonzalezDelgado2015}; \citealt{Belfiore2017}; \citealt{Goddard2017}). Measurements in these bands would then approximate global rather than local properties, undermining the motivation for a local analysis. We therefore restrict the local photometry to the optical surveys (at best $ugrizyY$), where the angular resolution permits genuinely local apertures. Although this limits the wavelength coverage of the local SEDs, it allows us to consistently estimate local stellar masses and rest-frame colours (Section~\ref{sec:excluding_phot}). Within the optical data, the worst-case PSF (Pan-STARRS $g$, FWHM 1.31$^{\prime\prime}$; \citealt{Magnier2020}) sets a minimum useful radius of $\sigma_{\rm PSF} = 0.56^{\prime\prime}$, corresponding to $\sim$1.5\,kpc at $z = 0.15$ (Figure~\ref{fig:ap_size_vs_z}), which we adopt as the physical radius of the local aperture. At lower redshift this aperture is well resolved in all optical bands; beyond $z = 0.15$ it falls below the PSF width, and local photometry is not performed.}

\subsubsection{Additional considerations}

Finally, to ensure consistent and reliable local photometry, we implement two quality-control checks within our pipeline. The first is to crossmatch each local aperture against the segmentation map and associated object masks for the image. This step is essential to verify that the measured flux is not contaminated by foreground stars, background galaxies, image artefacts, or neighbouring host-galaxy substructures that have been flagged during source detection and masking. If any portion of the local aperture overlaps with a masked region, the corresponding local photometric measurement is discarded. This conservative approach minimizes the risk of spurious flux contributions and ensures that retained measurements reflect uncontaminated light from the intended region of the host galaxy. This check removes 68 SNe (0.9\%) from the local photometry sample.

The second check ensures that the adopted local aperture remains physically meaningful relative to the size of the host galaxy. Specifically, we exclude local photometry for any system in which the projected physical radius of the local aperture exceeds the physical radius of the host galaxy. In such cases, the aperture would encompass a substantial fraction or entirety of the galaxy, and the resulting measurement would no longer represent a genuinely local environment but instead approximate a global quantity. This check removes a further 88 SNe (1.2\%). Imposing our redshift cut of $z\leq0.15$, 7377 (97.5\%) SNe have sufficiently well measured local photometry to proceed with SED fitting.

\begin{table*}
\centering
\caption{Model parameters and priors adopted in the \texttt{Bagpipes} SED fitting. $\mathcal{U}$ and $\mathcal{T}$ refer to uniform and Student's-$t$ distributions, respectively.}
\label{tab:bagpipes_params}
\begin{tabular}{llll}
\hline
Component & Symbol & Description & Prior / Value \\
\hline
Global & $z$ & Redshift & Fixed to best value (host or SN) \\
 & $t_\mathrm{bc}$ & Time-scale of birth clouds & $10$\,Myr \\
\hline
Non-parametric & $\log_{10}(M_*/M_\odot)$ & Total stellar mass formed & $\mathcal{U}(5, 13)$ \\
SFH & $\log_{10}(Z_*/Z_\odot)$ & Stellar metallicity & $\mathcal{U}(-2, 0.19)$ \\
 & $N_\mathrm{bins}$ & Number of star-formation bins & 7 \\
 & $t_\mathrm{bins}$ & Time bins of star formation in lookback time & $1^\mathrm{st}$ bin: [0, 30]\,Myr; \\
 & & & $2^\mathrm{nd}$ bin: [30, 100]\,Myr; \\
 & & & Final bin: [$0.9\,t_\mathrm{univ}(z),\ t_\mathrm{univ}(z)$] \\
 & & & Other bins linearly spaced$^{a}$ \\
 & $\Delta(\log_{10}\mathrm{SFR})$ & Change in SFR between adjacent bins & $\mathcal{T}(\mu=0;\ \sigma=0.3;\ \nu=2)$ \\
\hline
Nebular & $\log_{10} U$ & Ionization parameter & $\mathcal{U}(-4, -2)$ \\
\hline
Dust & Type & Salim & -- \\
 & $A_V$\,(mag) & Total $V$-band attenuation & $\mathcal{U}(0, 5)$ \\
 & $\epsilon$ & Scaling of $A_V$ for stars in birth clouds & 2.0 \\
 & $Q_\mathrm{PAH}$ & PAH mass fraction & $\mathcal{U}(0.1, 4.58)$ \\
 & $U_\mathrm{min}$ & Minimum radiation field & $\mathcal{U}(0.1, 25)$ \\
 & $\gamma_e$ & Fraction of starlight at $U_\mathrm{min}$ & $\mathcal{U}(0.0005, 1)$ \\
 & $\delta$ & Deviation of dust law slope from \citet{Calzetti2000} & $\mathcal{U}(-0.7, 0.3)$ \\
 & $B$ & Strength of 2175\,\AA\ bump & $\mathcal{U}(0, 5)$ \\
\hline
\end{tabular}
\vspace{2mm}
\begin{minipage}{\textwidth}
\footnotesize
$^{a}$\,Time bins are spaced linearly in lookback time rather than logarithmically, unlike the default implementation of \citet{Leja2019}. We find that this choice has negligible impact on the inferred stellar masses, star-formation rates, dust attenuations, and relative age ranking of galaxies. In our tests, the absolute values of the mass-weighted ages shift modestly, but galaxies inferred to be relatively young or old remain so. Some shift in the absolute calibration of mass-weighted age is expected, however, because age estimates are sensitive to the adopted SFH prior and temporal binning scheme \citep[e.g.,][]{Carnall2019, Wan2024}.
\end{minipage}
\end{table*}

\section{Modelling of Spectral Energy Distributions}\label{sec:sed_fitting}

\begin{figure*}
    \centering
    \includegraphics[width=\textwidth]{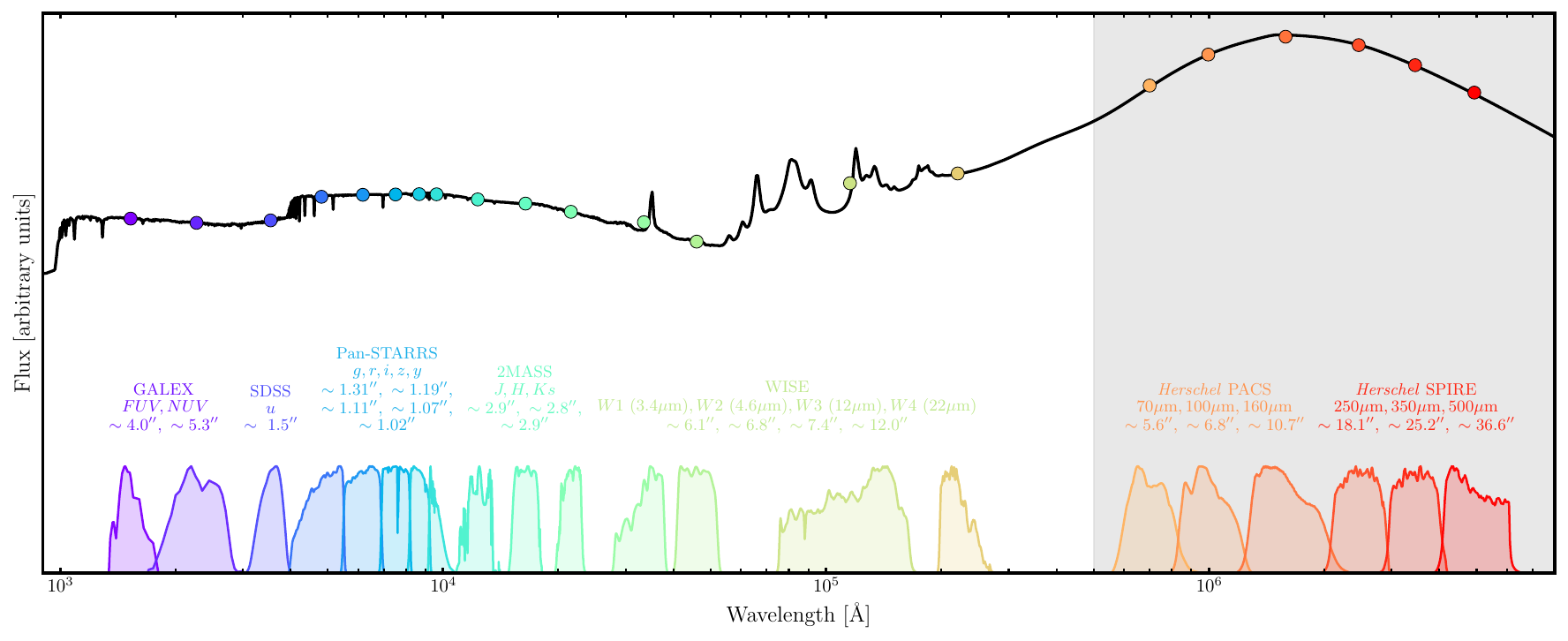}
    \caption{Simulated spectral energy distribution (SED) of a galaxy at $z=0.062$, illustrating the multiwavelength photometry used in this work, spanning the UV to far-infrared. Coloured points show the simulated photometry, while the black curve represents the underlying \texttt{Bagpipes} model. Filter transmission curves for each survey, and their associated median PSF FWHMs, are shown along the bottom, including \textit{GALEX} (FUV, NUV), SDSS ($u$), Pan-STARRS ($g,r,i,z,y$), 2MASS ($J,H,K_s$), \textit{WISE} (W1–W4), and \textit{Herschel} PACS and SPIRE bands. Transmission curves for DES ($g,r,i,z,Y$), SDSS ($g,r,i,z$), BASS ($g,r$), and MZLS ($z$) are omitted for clarity, as they largely overlap with the filters shown. The shaded region highlights the FIR regime, which is used in this work only for validation tests and is not typically included in the SED coverage for the majority of TITAN galaxies.}
    \label{fig:example_spectrum}
\end{figure*}

\begin{figure*}
    \centering
    \includegraphics[width=\textwidth]{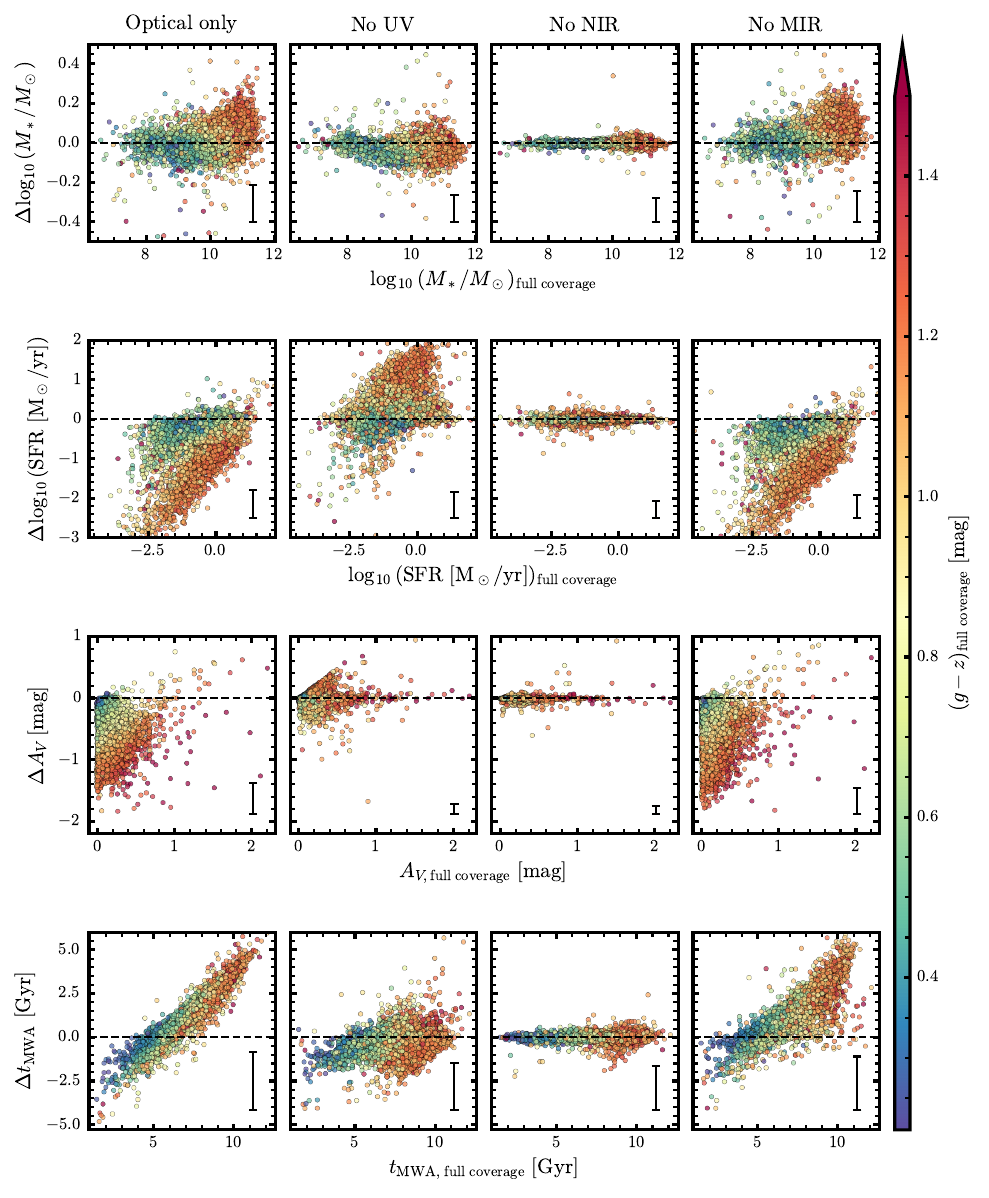}
    \caption{\rev{Offsets in the global host-galaxy properties fitted with \texttt{Bagpipes} when subsets of the photometry are excluded, $\Delta X = X_{\rm full} - X_{\rm restricted}$, as a function of the value derived with full coverage (measurements in all GALEX, 2MASS and WISE bands, and at least one set of optical bands). From left to right, the columns show fits using \revtwo{optical data only, and fits excluding the UV, the NIR, and the MIR}. From top to bottom, the rows show stellar mass ($M_*$), SFR, dust attenuation ($A_V$) and mass-weighted age ($t_{\rm MWA}$). Points are coloured by the full-coverage rest-frame $g-z$ colour, error bars show the median statistical uncertainty of the difference, and horizontal dashed lines mark zero offset.}}
    \label{fig:effect_of_missing_photometry}
\end{figure*}

For our host-galaxy analysis, we adopt the Bayesian Analysis of Galaxies for Physical Inference and Parameter EStimation (\texttt{Bagpipes}; \citealt{Bagpipes}) framework. \texttt{Bagpipes} constructs model galaxy SEDs by forward-modelling the emission from stellar populations, nebular gas, and dust. Stellar emission is generated by assuming a \cite{Kroupa2001} initial mass function (IMF), and integrating simple stellar populations from the 2016 revision of the \citet{Bruzual2003} stellar population synthesis models over a chosen star-formation history (SFH), which can be either parametric (e.g., exponentially declining, delayed-$\tau$, or double power-law) or non-parametric. Nebular emission lines and continuum are added using pre-computed grids of \texttt{CLOUDY}-based photoionization models \citep{Cloudy}, scaled by the ionizing photon output of the stellar component. Dust attenuation can be modelled using several prescriptions, including the two-component \citet{Charlot2000} model, the \citet{Calzetti2000} starburst attenuation law, and power-law or flexible broken-power-law formulations, allowing for differential obscuration between young and old stellar populations. The resulting rest-frame SED is then redshifted, attenuated by intergalactic medium absorption, and convolved with instrumental filter curves or spectral resolutions to enable direct comparison with observed photometric or spectroscopic data within a Bayesian inference framework.

\subsection{Fiducial modelling} \label{sec:fiducial_sed_fitting}

For our fiducial model, we adopt a non-parametric SFH because flexible forms demonstrably reduce systematic biases in key galaxy properties while better accommodating the diversity of SFH shapes. In controlled tests against hydrodynamic simulations, non-parametric models recover stellar masses with far smaller bias than common parametric forms, and also yield more accurate star-formation rates (SFRs) and ages \citep[e.g.,][]{Leja2017,Leja2019,Lower2020}. Parametric SFHs can often impose strong, implicit priors on SFR and age that can bias inferences and complicate mass-assembly studies \citep{Carnall2019}. Non-parametric models are not assumption-free, however, with results sensitive to the form of the SFH prior; we hence adopt well-motivated regularizing continuity priors \citep{Leja2019}.

For dust attenuation, we adopt the two-component flexible attenuation model implemented in \texttt{Bagpipes}, based on the formalism of \citet{Salim2018}. This model parametrizes the attenuation curve as a modified Calzetti-like law \citep{Calzetti2000} with a variable power-law slope and optional UV bump at 2175~\AA, allowing the shape of the attenuation curve to vary continuously rather than assuming a fixed form. The \citet{Salim2018} model provides sufficient freedom to capture observed diversity in attenuation slopes while remaining computationally efficient and well constrained for broadband photometry, making it well-suited to our dataset which encapsulates a diverse population of galaxies.

We further impose doubly enhanced attenuation for stars residing in birth clouds (with lifetimes up to $10\ \mathrm{Myr}$), relative to the diffuse interstellar medium. We also fit for the nebular ionization parameter $U$ and the parameters of the \citet{Draine2007} dust emission model of PAHs. Redshifts are fixed to their best available value ($z_\mathrm{host, spec}$ where available, otherwise $z_\mathrm{SN}$). Of the galaxies entering the SED fitting stage, \rev{all 7568 global photometric fits successfully converge}, alongside 7329 (99.3\%) of local photometric fits. \rev{The local fits that fail to converge have faint, low signal-to-noise local photometry (median S/N of 1.4 across bands, compared with 7.8 for the converged fits), and predominantly correspond to very low-mass hosts.} The adopted \texttt{Bagpipes} model is detailed in full in Table~\ref{tab:bagpipes_params}, and Figure~\ref{fig:example_spectrum} presents an example SED together with the filter transmission curves used in this work.

\subsection{The importance of broad photometric coverage}

In this section, we test how the inferred host-galaxy properties change when parts of the observed SED are removed from the fit. Restricting the wavelength coverage increases degeneracies between dust attenuation, stellar age, and recent star formation, and can bias the recovered parameters. Our goal here is therefore to identify which host-galaxy properties can be reported reliably for all of TITAN, and which should only be quoted when the SED includes the \rev{relevant bands} needed to break these degeneracies.

\subsubsection{Excluding photometric data} \label{sec:excluding_phot}

Motivated by the analysis of \Ramaiya, we repeat \rev{analogous tests} for TITAN by refitting the host-galaxy global photometry using \rev{multiple subsets of the available photometry: optical data only; excluding the NIR and MIR together; excluding the UV, NIR or MIR individually; excluding the GALEX NUV or FUV band; and excluding the WISE W3 or W4 band, individually or together}. We then compare these restricted fits to our fits with full coverage \rev{(measurements in all GALEX, 2MASS and WISE bands, and for at least one set of optical bands; Table~\ref{tab:coverage})}, which we take as the most comprehensive baseline for the low-redshift TITAN sample. \revtwo{The results of the tests excluding entire wavelength regimes are presented in Figure~\ref{fig:effect_of_missing_photometry}, and those excluding individual bands or combinations of regimes in Appendix~\ref{app:excl_phot} (Figure~\ref{fig:excl_phot_app})}, where $\Delta X = X_{\rm full} - X_{\rm restricted}$.

\rev{We first note that excluding the 2MASS NIR photometry has no appreciable effect on any of the inferred properties (\revtwo{third} column of Figure~\ref{fig:effect_of_missing_photometry}), and that fits excluding both the NIR and MIR are nearly indistinguishable from those excluding the MIR alone. At $z \lesssim 0.15$, the $J$, $H$ and $K_s$ bands probe rest-frame wavelengths of $\sim$1--2\,$\mu$m, where the SED is dominated by the same evolved stellar populations already constrained by the reddest optical bands and WISE W1 and W2 \revtwo{\citep[e.g.,][]{Jarrett2013, Meidt2014}}, and the 2MASS photometry is comparatively shallow, with only 49--64\% of hosts detected (Table~\ref{tab:coverage}). \revtwo{\citet{Lee2026} similarly found that adding 2MASS photometry degraded their SED fits to SN~Ia host galaxies, while changing their stellar masses by only $\sim$0.01\,dex on average.} We therefore do not require NIR coverage for any property.}

The stellar masses are highly stable. Even for optical-only fits, the systematic bias remains small ($\sim0.01$--$0.02\,\mathrm{dex}$) and the scatter is modest ($\sigma_{\Delta M_*}\sim0.1\,\mathrm{dex}$)\rev{, although the reddest, most massive hosts show offsets of up to $\sim$0.3\,dex in the absence of MIR data}. This is expected for a low-redshift sample such as TITAN ($z \lesssim 0.15$): the reddest observed optical bands already probe rest-frame wavelengths near $\sim0.8$--$1.0\,\mu\mathrm{m}$, where the light is dominated by older stellar populations (G/K dwarfs and RGB stars) with relatively stable mass-to-light ratios \revtwo{\citep[e.g.,][]{Bell2001, Bell2003}}. As a result, even limited optical coverage retains most of the information needed to recover robust stellar masses.

In contrast, the inferred SFRs are much more sensitive to the removal of \rev{UV and} MIR data. When only UV--NIR information is available \revtwo{(fourth column of Figure~\ref{fig:effect_of_missing_photometry})}, the reddest and most weakly star-forming galaxies show systematically negative $\Delta\log_{10}(\mathrm{SFR})$, meaning that their SFRs are overestimated relative to the fiducial UV--MIR solution, by up to $\sim$2--3\,dex. A single measurement in either WISE W3 or W4, even a non-detection, largely mitigates this effect \revtwo{(Figure~\ref{fig:excl_phot_app})}, whereas W1 and W2 alone do not (see below). This behaviour is qualitatively consistent with \Ramaiya\ and reflects the well-known difficulty of distinguishing genuinely old stellar populations from dust-reddened, weakly star-forming ones when the dust-emission component is not directly constrained \revtwo{\citep[e.g.,][]{Papovich2001, Conroy2013}}. Conversely, when only optical--MIR information is available \revtwo{(second column of Figure~\ref{fig:effect_of_missing_photometry})}, the SFRs of red and intermediate-colour hosts are systematically underestimated, by up to $\sim$2\,dex. Although the MIR constrains the total dust emission, it does not constrain which stellar populations heat the dust: without the UV, the fit can instead reproduce the dust emission with a lower recent SFR and a larger contribution from dust heated by older stars\revtwo{, which can be substantial in nearby galaxies \citep{Nersesian2019}}. \revtwo{As shown in Figure~\ref{fig:excl_phot_app}, the} NUV provides the dominant constraint, but excluding the FUV alone still introduces residual trends of up to $\sim$1\,dex for the reddest hosts, while the FUV alone is not sufficiently constraining, as it is more strongly attenuated by dust \citep{Calzetti2000} and detected in a considerably smaller fraction of hosts (Table~\ref{tab:coverage}); both UV bands are therefore required (see also below).

The effect is even clearer for dust attenuation. In the third rows of \revtwo{both Figure~\ref{fig:effect_of_missing_photometry} and Figure~\ref{fig:excl_phot_app}}, excluding the MIR bands leads to systematically larger inferred $A_V$, \rev{by up to $\sim$2\,mag,} with the strongest offsets occurring for intrinsically dust-poor, optically red systems. In these cases, the fits compensate for the lack of direct dust-emission information by invoking additional attenuation to reproduce the observed colours. This in turn drives the corresponding SFR biases, since the models attribute a larger fraction of the red continuum to obscured star formation rather than to an older stellar population. \rev{As for the SFR, a measurement in either W3 or W4 largely removes this bias, whereas W1 and W2 alone are not sufficient. At $z \lesssim 0.15$, the W1 and W2 bands largely trace the stellar continuum, where attenuation is negligible and dusty and dust-free solutions predict similar fluxes, whereas the W3 and W4 bands trace dust emission \revtwo{\citep[e.g.,][]{Jarrett2013, Cluver2017}}, where the two solutions differ strongly: within the energy-balance framework, the attenuated light must be re-emitted at these wavelengths, so that even an upper limit from a non-detection rules out the heavily attenuated solutions. Excluding the UV instead affects a subset of moderately dusty hosts, whose attenuation collapses to $A_V \approx 0$ (shifts of up to $\sim$0.5\,mag, well in excess of the typical statistical uncertainty), together with SFRs lower by $\sim$0.9\,dex. These hosts are well detected in both GALEX bands but have red UV colours (median FUV$-$NUV $= 1.34$, compared with 0.66 for hosts of similar attenuation whose $A_V$ is unaffected), and are essentially undetected in W3 and W4. With both UV bands, the red UV slope requires a dust-reddened young stellar population \revtwo{\citep{Meurer1999}}; without them, the UV-to-optical light is instead attributed to a dust-free older population \revtwo{\citep{Kong2004}}, a solution permitted within the energy-balance framework only because the weak MIR emission places little constraint on the dust luminosity. A single UV band is not sufficient to remove this effect, as excluding either the NUV or the FUV alone produces a similar shift \revtwo{(see Figure~\ref{fig:excl_phot_app})}: it is the FUV$-$NUV colour that helps to constrain the reddening of the young population.}

The mass-weighted ages show \rev{similar} underlying behaviour. When the MIR bands are removed, galaxies that are old in the fiducial UV--MIR fits -- particularly red, quiescent systems -- are preferentially assigned younger ages, by up to $\sim$5\,Gyr, while the youngest hosts are assigned older ages. Without the MIR, the ages retain little information, clustering around $\sim$5\,Gyr regardless of the fiducial value\revtwo{, as expected when the age is largely set by the SFH prior \citep{Carnall2019, Leja2019}: the continuity prior we adopt favours smooth SFHs which correspond to intermediate mass-weighted ages.} This is the expected counterpart to the SFR and $A_V$ shifts. The addition of UV and NIR data \rev{does not appreciably mitigate this effect}, indicating that MIR photometry provides the key leverage needed to stabilize age estimates in systems where the optical colours alone are degenerate. \rev{A measurement in either W3 or W4 removes the bulk of this effect; with W1 and W2 alone, the ages of the oldest hosts remain underestimated by up to $\sim$1--2\,Gyr. Excluding the UV instead leads to systematically older ages for the youngest, bluest hosts, by $\sim$1--3\,Gyr, as the young stellar population is then unconstrained. Including a single UV band removes most of this bias, but a subset of red, older hosts is still assigned systematically older ages, by up to $\sim$2--3\,Gyr, so that, as for the SFR and $A_V$, both UV bands are required.}

The main conclusion is therefore straightforward: at the redshifts of the TITAN sample, stellar mass and rest-frame colour are robustly constrained by optical data alone, whereas reliable SFRs, dust attenuation and stellar ages require coverage in both GALEX bands, which together constrain the young stellar population and its attenuation, and in WISE W3 or W4, which constrain the dust emission and break the age--dust degeneracy. Coverage, rather than detection, is what matters, as non-detections in these bands still exclude the degenerate solutions. Of the 7568 hosts with successful global SED fits, 5842 (77.2\%) have a measurement in both GALEX bands and at least one of W3 or W4; only 24 hosts lack coverage in both W3 and W4. \revtwo{These conclusions are consistent with the wider galaxy SED-fitting literature, which has long shown that stellar masses are relatively insensitive to wavelength coverage when the rest-frame optical is sampled, whereas SFRs, dust attenuation and stellar ages require constraints on both the unobscured (UV) and obscured (IR) components \citep[e.g.][]{Walcher2011, Wuyts2011, Conroy2013, Chang2015, Salim2016, Hunt2019, Pacifici2023}.}

\rev{Finally, motivated by \citet{Clark2026}, who investigated active galactic nucleus (AGN) contamination in the ZTF SN~Ia sample, we consider the effect of AGN on our host properties. As our SED model does not include emission from an AGN, the mid-infrared emission of hosts with a luminous AGN is instead attributed to dust heated by star formation, biasing their inferred SFRs and $A_V$ high. We identify such hosts using the WISE colour criterion of \citet{Stern2012} (W1$-$W2\,$\geq$\,0.8 for W2\,$<$\,15.05, in Vega magnitudes), which selects AGN with $\sim$95\% reliability at this depth. This identifies 15 hosts (0.3\%) which have the coverage defined above. As expected, their fits imply unusually high levels of star formation (median $\log_{10}(\mathrm{sSFR/yr^{-1}}) = -9.3$, compared with $-10.1$ for the full sample), with three exceeding 50\,$M_\odot$\,yr$^{-1}$. Because our colours are measured within apertures enclosing the whole host, the stellar emission of the host dilutes that of the AGN, so this selection is conservative and identifies only AGN that dominate the mid-infrared emission of their host. Accordingly, in TITAN DR1 we report stellar masses and $g-z$ colours for all 7568 hosts with successful global SED fits, and SFRs, sSFRs, $A_V$ and $t_{\rm MWA}$ for the 5827 (77.0\%) hosts with the required coverage and no evidence of a luminous AGN. As only optical data are considered for the local photometry, the only local properties we report are stellar mass and $g-z$ colour, which are included for all 7329 converged local SED fits.}

\subsubsection{Including FIR data}

\begin{figure*}
    \centering
    \includegraphics[width=\textwidth]{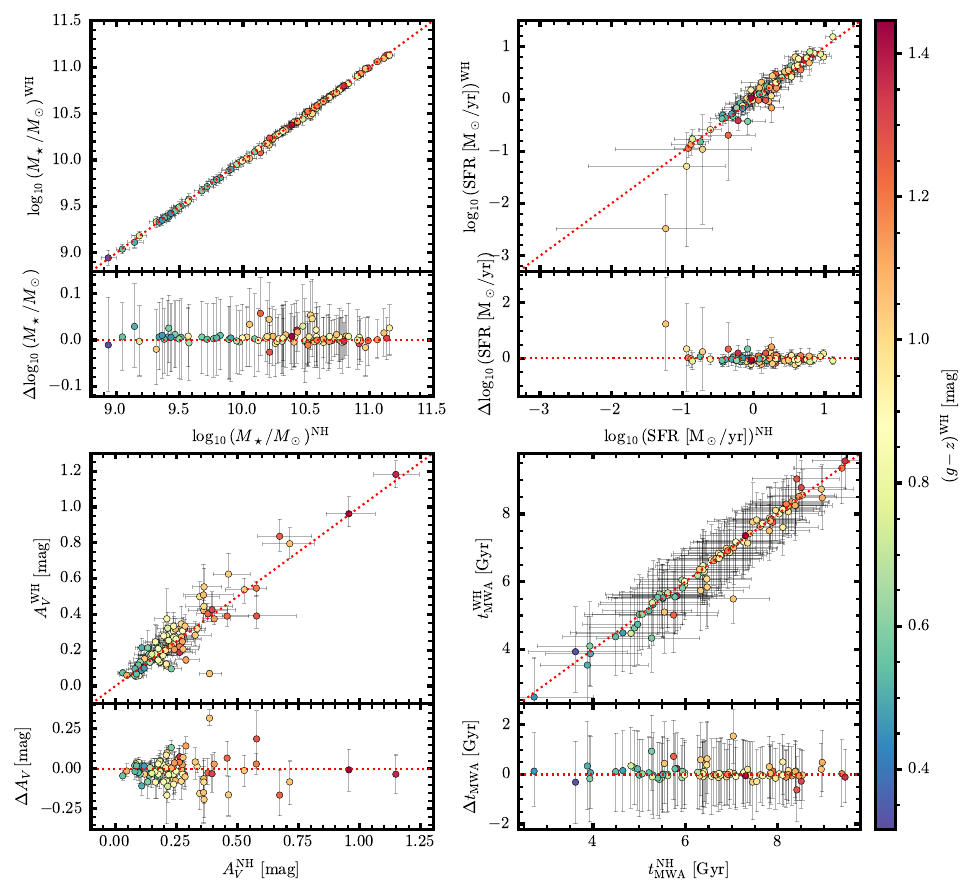}
    \caption{Comparison between fitted \texttt{Bagpipes} global galaxy properties derived either with (WH) or without (NH) the inclusion of FIR data from the \textit{Herschel} Space Observatory. The upper panels show, from top-left to bottom-right, relationships between stellar mass ($M_*$), SFR, dust attenuation ($A_V$), and galaxy mass-weighted age ($t_\mathrm{MWA}$). The lower panels show the corresponding residuals, shown as $X^\mathrm{NH}-X^\mathrm{WH}$, where $X$ represents any of the above four quantities. Points are coloured according to their $g-z$ colour (derived with FIR data), and red, dotted lines show the one-to-one line (upper panels) or the zero-residual line (lower panels).}
    \label{fig:herschel_comp}
\end{figure*}

While all-sky MIR data are readily available, there remains a notable lack of wide-field FIR coverage. For TITAN, the practical question is whether UV--MIR coverage is already sufficient at low redshift, or whether the absence of FIR data could still bias the host-galaxy properties we report. Because the TITAN hosts lie in the nearby Universe, the observed MIR bands probe a larger fraction of the dust-emission SED than they do at higher redshift, so FIR information may be less critical here than in samples such as DES5YR. In this section, we test that directly by comparing fits with and without available \textit{Herschel} photometry.

We compile FIR data from the \textit{Herschel} Space Observatory, utilizing observations from the Photodetector Array Camera and Spectrometer (PACS; $70$, $100$, and $160\mu$m) and the Spectral and Photometric Imaging Receiver (SPIRE; $250$, $350$, and $500\mu$m). We make use of galaxy photometry from the DMU32 All-Sky `A-List' catalogue of the \textit{Herschel} Extragalactic Legacy Project (HELP; \citealt{Shirley2021}). FIR fluxes in this catalogue are derived using \texttt{XID+} \citep{Hurley2017}, a Bayesian prior-based de-blending tool that forward-models the PACS and SPIRE maps using source positions from the HELP master list as priors, convolved with the instrumental PSF (FWHM of $\sim5.6, \sim6.8, \sim10.7^{\prime\prime}$ at 70, 100, 160~$\mu\mathrm{m}$ \citep{Balog2014} and $\sim18.1, \sim25.2, \sim36.6^{\prime\prime}$ at 250, 350, 500~$\mu\mathrm{m}$ \citep{Swinyard2010}). The `A-List' comprises galaxies drawn from all \textit{Herschel} survey fields that have at least \rev{two optical, two NIR and two FIR detections (S/N$>2$), a photometric / spectroscopic redshift}, and a successful SED fit. We identify \rev{98} TITAN SN Ia host galaxies within this catalogue.

To isolate the impact of FIR data alone when comparing SED fits with and without \textit{Herschel} coverage, we additionally require \rev{full photometric coverage as outlined in Section~\ref{sec:excluding_phot}}. This ensures that any differences in the inferred physical parameters arise solely from the inclusion of FIR photometry. Applying this requirement reduces the sample considered in this section to \rev{92} galaxies, \rev{spanning $0.02 \leq z \leq 0.12$, $0.3 \leq (g-z)\ [\mathrm{mag}] \leq 1.5$, $9.0\leq \log_{10}(M_*/M_\odot)\leq11.2$. The subsample is largely representative of the TITAN sample as a whole, with a slight bias towards high-mass, star-forming hosts (median stellar mass $\log_{10}(M_*/M_\odot) = 10.38$ and 75\% star-forming, compared with 10.22 and 71\% for the full sample; see Sections~\ref{sec:global_mass} and \ref{sec:global_sfr}), due to the requirement of multiple FIR detections.}

We fit the photometric data using identical methodology to that described in Section~\ref{sec:fiducial_sed_fitting} and Table~\ref{tab:bagpipes_params}. The resulting comparison between parameters derived without (NH) and with \textit{Herschel} data (WH) is shown in Figure~\ref{fig:herschel_comp}. We find excellent agreement between the two sets of derived properties. \rev{Defining $\Delta X = X^{\rm NH} - X^{\rm WH}$, the median offsets and scatter (nMAD) are $+0.003$ and 0.007\,dex in stellar mass, $-0.026$ and 0.09\,dex in SFR, and $+0.007$ and 0.13\,Gyr in $t_{\rm MWA}$, in each case well within the typical statistical uncertainties of the fiducial fits (0.04\,dex, 0.14\,dex, and 0.9\,Gyr, respectively).} Importantly, this consistency persists across the full range of $g-z$ colour, stellar mass, SFR, $A_V$, and $t_\mathrm{MWA}$. 

\rev{The largest differences are observed in $A_V$, for which the inclusion of FIR data increases the attenuation slightly on average (median $\Delta A_V = -0.016$\,mag, nMAD 0.05\,mag), comparable to the typical uncertainty of 0.045\,mag. Differences in $A_V$ are somewhat expected; the WISE W3 and W4 bands trace PAH emission and warm dust near star-forming regions, which constitute only a fraction of the total dust luminosity, so that without FIR data the absolute dust luminosity, and hence the precise attenuation, is less tightly constrained.}

Overall, our results demonstrate that MIR constraints alone are sufficient to mitigate the age--dust degeneracy at $z \lesssim 0.15$. We therefore conclude that the host-galaxy properties of the TITAN sample -- derived in the absence of FIR data, but including UV--MIR photometry -- are robust, and suitable for subsequent analyses exploring correlations between SN~Ia properties and those of their host galaxies beyond mass and colour alone.

For the next generation of wide-field high-$z$ SN surveys, where all-sky FIR data will remain scarce, radio-calibrated SFRs offer a robust alternative approach to break the age--dust degeneracy \citep{Ramaiya2026}.

\section{Derived Properties and Sample Statistics} \label{sec:sample_props}

In this section, we present the distributions of the physical galaxy properties of the TITAN SN Ia sample. We first present the global properties, followed by the local properties, and draw comparisons to other samples in the literature throughout, where possible.

\subsection{Global properties}

\subsubsection{Stellar mass} \label{sec:global_mass}

\begin{figure}
    \centering
    \includegraphics[width=\columnwidth]{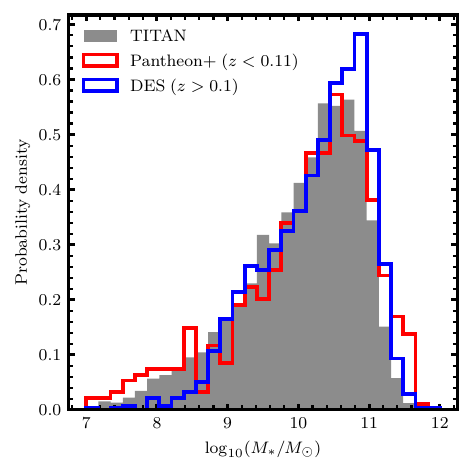}
    \caption{Host-galaxy stellar mass distributions for the TITAN SN~Ia sample compared to the Pantheon+ low-$z$ ($z<0.11$) and DES-SN5YR high-$z$ ($z>0.1$) samples. The TITAN distribution is shown as the filled grey histogram, while the comparison samples are overlaid as outlined histograms.}
    \label{fig:host_mass_histogram}
\end{figure}

\begin{figure*}
    \centering
    \includegraphics[width=\textwidth]{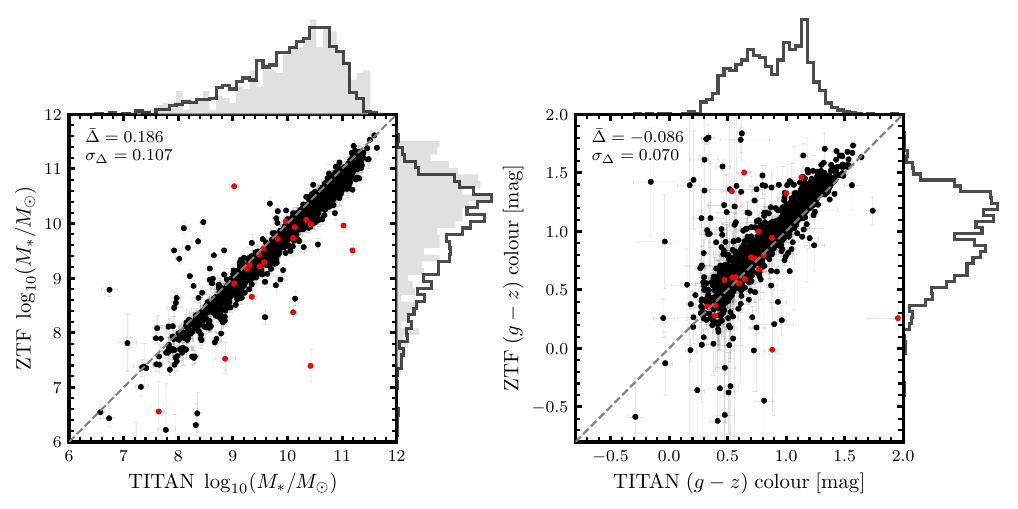}
    \caption{Comparison of host-galaxy properties derived by TITAN and ZTF for the \rev{1947} SNe Ia common to both samples with successful SED fits. In each panel, the grey dashed line marks the one-to-one relation, and SNe whose TITAN and ZTF host-galaxy coordinates disagree by more than $2^{\prime\prime}$ are shown in red. The values quoted in the upper-left of each panel are the mean offset $\bar{\Delta}$ and sigma-clipped scatter $\sigma_\Delta$, with $\Delta \equiv x_\mathrm{TITAN} - x_\mathrm{ZTF}$. \textit{Left:} Global stellar mass. The mean offset is $\bar{\Delta} = \rev{0.186}$~dex (ZTF less massive on average), with $\sigma_\Delta = \rev{0.107}$~dex. The top and right marginal histograms show the TITAN and ZTF mass distributions respectively, with the Pantheon+ low-$z$ sample overlaid in grey for comparison. \textit{Right:} Global $(g-z)$ colour. The mean offset is $\bar{\Delta} = \rev{-0.086}$~mag (ZTF redder on average), with $\sigma_\Delta = \rev{0.070}$~mag. The marginal histograms show the corresponding $(g-z)$ distributions.}
    \label{fig:titan_ztf_comparison}
\end{figure*} 

The distribution of TITAN global host galaxy stellar masses is shown in Figure~\ref{fig:host_mass_histogram}, alongside those of the Pantheon+ low-$z$ sample (\citealt{Scolnic2022}; $z \lesssim 0.11$) and the DES-SN5YR high-$z$ sample (\citealt{DES_Dovekie}; $z \gtrsim 0.1$) for comparison.

Adopting the conventional stellar-mass division at $\log_{10}(M_*/M_\odot)=10$, we find that \rev{$60\%$} of TITAN SNe~Ia in the bronze sample occur in high-mass galaxies, with a median value of $\log_{10}(M_*/M_\odot)=10.22$. 
For comparison, the high-mass fractions are $63\%$ for the Pantheon+ low-$z$ sample and $67\%$ for the DES-SN5YR high-$z$ sample, with correspondingly higher medians of 10.31 and 10.39, respectively. Overall, the TITAN host-mass distribution is in good agreement with these literature samples.

The slightly elevated high-mass fraction in Pantheon+ low-$z$ can be understood in the context of its constituent surveys, many of which were galaxy-targeted searches that monitored pre-selected, luminous nearby galaxies. Such strategies preferentially discover SNe in massive hosts, producing the prominent high-mass shoulder at $\log_{10}(M_*/M_\odot) \gtrsim 11$ seen in the distribution. 
The untargeted nature of the ATLAS survey \citep[see][for a description of SN discovery in ATLAS]{SmithATLAS2020} combined with the magnitude-limited follow-up of all the contributing spectroscopic classification and follow-up programmes (see Section\,\ref{sec:sn_z}) results in a more representative sampling of the low-redshift galaxy population in TITAN.

Similarly, in high-$z$, narrow field, surveys such as DES, selection effects also tend to favour the selection of cosmologically useful SNe Ia residing in more massive host galaxies. Low-mass, low-surface-brightness hosts are more difficult to detect and to quantitatively observe with optical spectra particularly as redshift increases. Spectroscopic follow-up strategies often prioritize brighter hosts to maximize redshift completeness and signal-to-noise; for example, DES actively followed-up all SNe in hosts with $m_r<24\ \mathrm{mag}$ \citep{Lidman2020}. These factors collectively skew high-$z$ samples towards more massive galaxies, as demonstrated by the lack of a low-mass tail below $\log_{10}(M_*/M_\odot) \lesssim 8$ in the DES data in Figure \ref{fig:host_mass_histogram}.

The TITAN sample also contains a significant overlap with the ZTF DR2 dataset \citep{Rigault2025}; we find that when considering normal or 91T-like SNe Ia, there exists an overlap of \rev{1947} objects with a successful global mass measurement of the identified host galaxy. This provides a useful opportunity to compare the two datasets, particularly their matched galaxies and mass-measurement methodology. A comparison of their mass distributions is presented in the left panel of Figure \ref{fig:titan_ztf_comparison}.

We find that $\sim99\%$ of TITAN and ZTF SNe Ia are assigned to the same host galaxy (within a $2^{\prime\prime}$ tolerance), and find that the measured masses between the two samples are extremely consistent, with a scatter of only \rev{$0.107\ \mathrm{dex}$}. We also find a systematic offset in the reported masses, with the masses reported by ZTF systematically lower than those of TITAN by \rev{$0.186\ \mathrm{dex}$}. As a consequence, approximately \rev{$9\%$} of objects switch sides of the canonical host-mass step boundary at $\log_{10}(M_*/M_\odot)=10$.

Even when adopting the same \cite{Kroupa2001} IMF, literature comparisons consistently find 0.1–0.3 dex systematic differences between independent SED-fitting pipelines applied to the same photometry \citep[e.g.,][]{Conroy2009, Conroy2013}. The offset observed here is therefore entirely consistent with methodological differences between PEGASÉ.2-based \citep{Fioc1997, LeBorgne2002} (as used for ZTF host galaxy SED fitting; \citealt{Rigault2025}) and \texttt{Bagpipes}-based mass estimates \citep[which use the 2016 revision of the stellar population synthesis of][]{Bruzual2003}
rather than reflecting a physical discrepancy between the host galaxy samples.

\subsubsection{$g-z$ colour}

\begin{figure}
    \centering
    \includegraphics[width=\columnwidth]{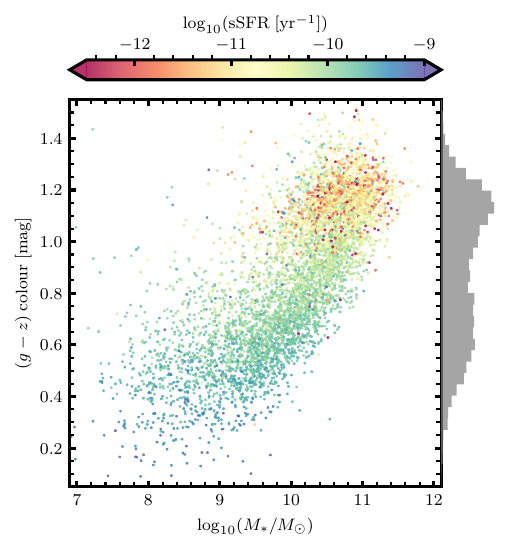}
    \caption{The distribution of global $g-z$ colour as a function of global stellar mass, coloured by global sSFR.}
    \label{fig:gz_colour}
\end{figure}

\begin{figure*}
    \centering
    \includegraphics[width=\textwidth]{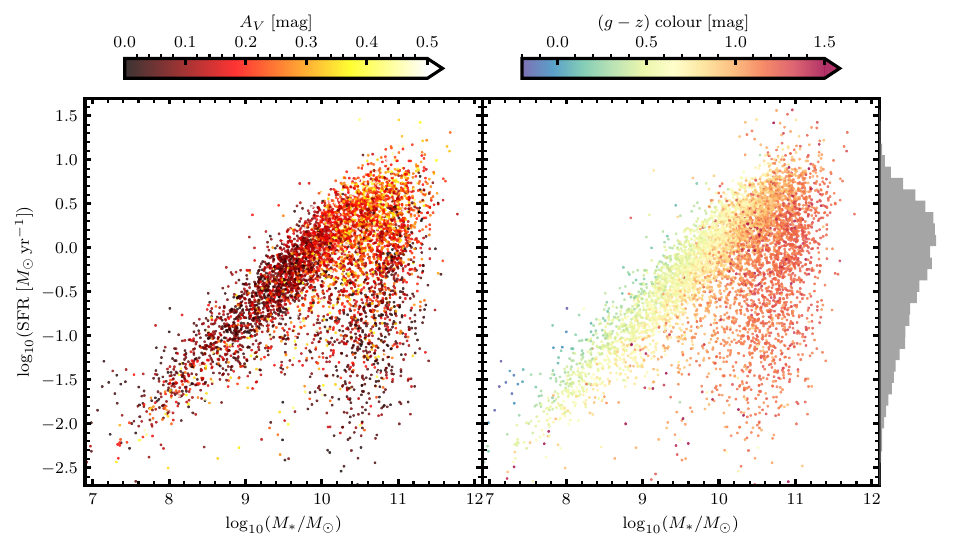}
    \caption{The distribution of global SFR as a function of global stellar mass, coloured by global $A_V$ (left) and global $g-z$ colour (right).}
    \label{fig:sfms}
\end{figure*} 

The distribution of $g-z$ colour for the TITAN host galaxies is shown in Figure \ref{fig:gz_colour}. The sample spans a broad, clearly bimodal range in colour, extending from blue-cloud systems associated with actively star-forming galaxies to red-sequence systems dominated by older, quiescent stellar populations, with a substantial intermediate population occupying the green valley. This bimodality is a well-established feature of the low-redshift galaxy population \citep[e.g.,][]{Strateva2001, Baldry2004}, and reflects the diversity of galaxy evolutionary states represented within the TITAN host population.

Comparing the rest-frame colour distributions in the right panel of Figure~\ref{fig:titan_ztf_comparison}, we observe broad consistency between the two samples, with sigma-clipped scatter \rev{$\sigma_\Delta = 0.070$~mag}, but a systematic offset of \rev{$\bar{\Delta} = -0.086$~mag}, indicating that ZTF colours are on average redder than those in TITAN. Offsets of this magnitude between SED-fitting pipelines are, again, not unusual \citep[e.g.,][]{Conroy2013, Pacifici2023}, and could also arise
from differences in the host-photometry methods adopted by each catalogue. 

Rest-frame $g-z$ colour cannot uniquely determine a galaxy's physical state -- Figure~\ref{fig:sfms} makes this clear, with hosts at fixed $g-z$ spanning more than three orders of magnitude in inferred SFR. What colour does offer is a low-dimensional classifier that is relatively prior independent when there is a dearth of data; properties such as the instantaneous SFR or stellar age are much more dependent on the chosen SED model. In this context, the TITAN sample is especially valuable. Its breadth and depth permit a direct comparison of colour-based, mass-based, and more model-dependent standardization schemes within a single, homogeneous low-redshift dataset, against which the model dependence of the latter can be calibrated. Because optical colours remain measurable at higher redshift where multi-wavelength coverage often degrades, they also provide a tractable proxy for separating hosts by evolutionary state across the full SN Ia redshift range -- with the caveat that their quantitative interpretation is limited by precisely the scatter visible in Figure~\ref{fig:sfms}.

\subsubsection{Star formation rate} \label{sec:global_sfr}

We derive global SFRs and specific SFRs (sSFR) from the fitted \texttt{Bagpipes} SFHs, and report values averaged over the most recent $100\ \mathrm{Myr}$ of each galaxy's history. The resulting distribution is shown in Figure \ref{fig:sfms}. 

The TITAN sample spans a wide dynamic range in star-formation activity, from quiescent galaxies with very low inferred SFRs to actively star-forming systems, with median values of \rev{$\log_{10}(\mathrm{SFR} [M_\odot/\mathrm{yr}])=-0.16$} and \rev{$\log_{10}(\mathrm{sSFR} [\mathrm{yr^{-1}}])=-10.12$}. Adopting standard literature thresholds of $\log_{10}(\mathrm{sSFR}/\mathrm{yr}^{-1}) > -10.5$ for star-forming galaxies and $\log_{10}(\mathrm{sSFR}/\mathrm{yr}^{-1}) < -11.0$ for quiescent systems, we find that the sample is predominantly composed of star-forming hosts, which account for \rev{71\%} of the population. In contrast, only \rev{16\%} of galaxies are classified as quiescent, with the remaining \rev{13\%} occupying an intermediate, transitional regime.

In the SFR-$M_*$ plane, the host galaxies trace the familiar structure of the galaxy population, including a well-defined star-forming main sequence in which SFR increases with stellar mass, as well as a distinct population of massive, low-SFR systems. This bimodal structure reflects the underlying division between star-forming disc galaxies and quiescent, elliptical systems, and demonstrates that the TITAN hosts are broadly representative of the general low-redshift galaxy population. The observed trends in dust attenuation and colour across this plane are also consistent with expectations from galaxy evolution studies: galaxies with higher SFRs tend to exhibit bluer colours and higher dust attenuation, while quiescent systems occupy a region of low SFR, red colour, and low dust content.

This overall structure mirrors the three-region division presented by \Ramaiya and supports the interpretation that SNe Ia probe host galaxies at different evolutionary stages. \rev{This dataset therefore provides a natural framework to subdivide the SN~Ia population by host location in the SFR--$M_*$ plane. Several analyses have found that the SN~Ia colour--luminosity relation, or the effective dust law it reflects, differs between host environments, whether defined by global or local host properties \citep{BroutScolnic2021, GonzalezGaitan2021, Kelsey2021, Kelsey2023}; TITAN enables such environment-dependent colour standardization to be tested in a single large, homogeneous low-redshift sample, which we defer to future work.}

\subsubsection{Dust attenuation}

\begin{figure}
    \centering
    \includegraphics[width=\columnwidth]{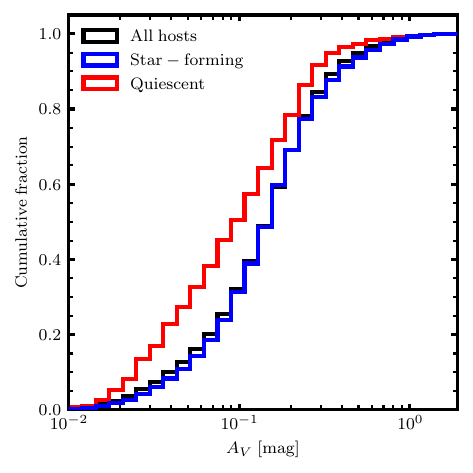}
    \caption{Cumulative distribution function of global $A_V$, shown both for the sample as a whole and for subsamples of star-forming ($\log_{10}(\mathrm{sSFR}/\mathrm{yr}^{-1}) > -10.5$) and quiescent ($\log_{10}(\mathrm{sSFR}/\mathrm{yr}^{-1}) < -11$) galaxies.}
    \label{fig:av}
\end{figure}

The distribution of V-band dust attenuation ($A_V$) for the TITAN host galaxies is shown in Figure \ref{fig:av}. The distribution is strongly peaked at low attenuation values, with the majority of galaxies exhibiting modest dust content and a tail extending towards more heavily obscured systems; the median $A_V$ for the sample is 0.14~mag. The observed behaviour is consistent with the mixed nature of the host population. Quiescent, massive galaxies typically require little dust attenuation to reproduce their red colours, while actively star-forming systems populate the higher-$A_V$ tail, reflecting the presence of dust associated with ongoing star formation.

The observed $A_V$ distribution is approximately exponential, consistent with the attenuation distributions commonly adopted in SN~Ia scatter models. Most contemporary models \citep[e.g.,][]{BroutScolnic2021,Thorp2021,Popovic2023,Wojtak2025} assume an exponential prior on $A_V$. The empirical similarity between the TITAN host-galaxy distribution and these assumed priors provides observational support for the functional forms typically used in SN~Ia population modelling.

The distinction between global attenuation and line-of-sight extinction is particularly important in the context of SN Ia standardization. Several recent studies have argued that host-galaxy dust, or variations in the effective dust law, may contribute to the observed correlations between SN Ia Hubble residuals, colour, and host-galaxy properties, including the stellar-mass step \citep[e.g.,][]{BroutScolnic2021, Popovic2023, Kelsey2023, Duarte2023}. \rev{However, analyses combining optical--NIR SN light curves with host-galaxy attenuation measurements suggest that dust alone may not fully explain these trends \citep[e.g.,][]{Thorp2022, Duarte2023, Popovic2024, Grayling2024, Peterson2024}. Moreover, SED-derived attenuation is a galaxy-averaged quantity, whereas the extinction towards an individual SN depends on its local environment and the viewing geometry. The $A_V$ values presented here therefore describe the integrated interstellar medium of each host, rather than the immediate environment of the progenitor\revtwo{; indeed, \citet{Kelsey2026} finds that $A_V$ shows the largest discrepancy between the local (3\,kpc) and global properties of nearby SN~Ia hosts}. They nevertheless provide a basis for testing whether host attenuation carries information beyond stellar mass, for example through correlations with Hubble residuals, colour-law parameters or intrinsic-scatter terms.}

\subsubsection{Mass-weighted age}

\begin{figure*}
    \centering
    \includegraphics[width=\textwidth]{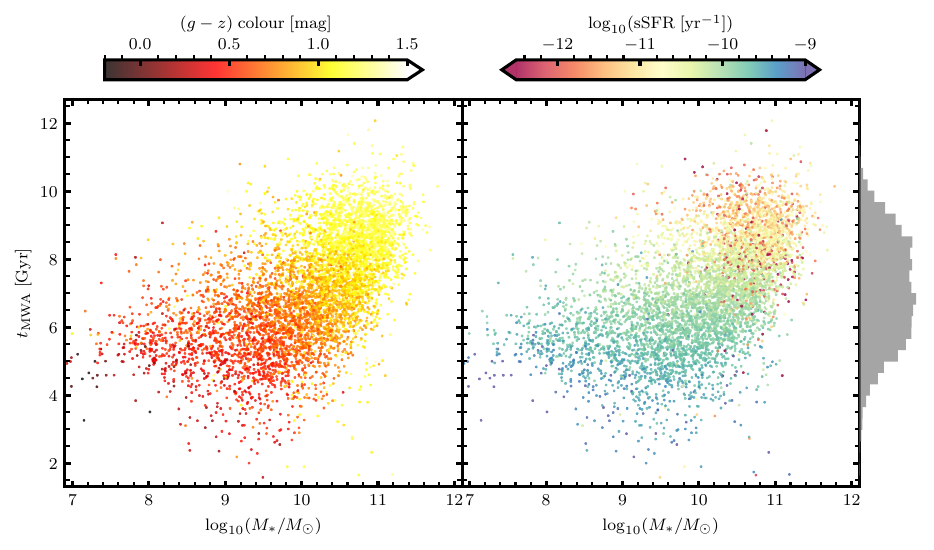}
    \caption{Global galaxy mass-weighted age as a function of global stellar mass, coloured by global $g-z$ colour (left) and global sSFR (right).}
    \label{fig:age}
\end{figure*} 

\begin{figure*}
    \centering
    \includegraphics[width=\textwidth]{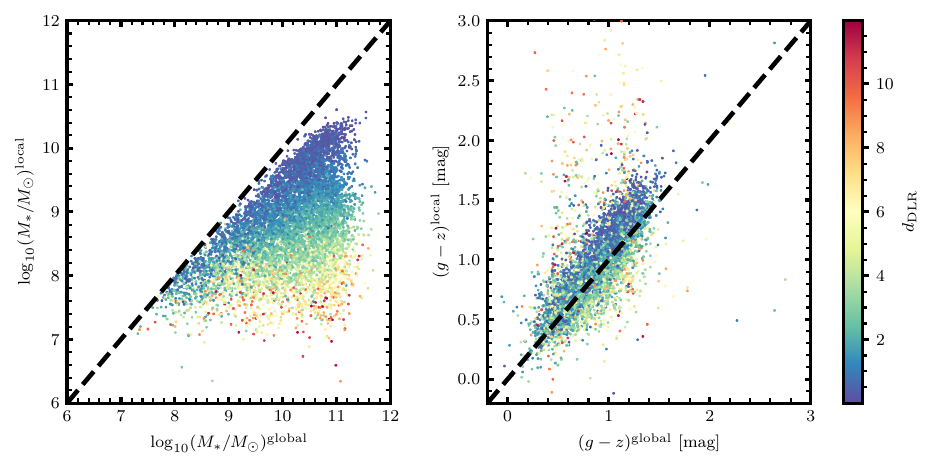}
    \caption{Comparison between global and local host-galaxy properties. In both panels, points are coloured according to $d_\mathrm{DLR}$, and the dashed black line indicates the one-to-one relation. \textit{Left:} local stellar mass, measured within a fixed 1.5~kpc aperture centred on the SN position, as a function of the corresponding global host-galaxy stellar mass. \textit{Right:} local rest-frame $g-z$ colour as a function of global host-galaxy $g-z$ colour.}
    \label{fig:local_props}
\end{figure*}

Mass-weighted stellar population age ($t_\mathrm{MWA}$) provides a physically intuitive measure of the average formation time of a galaxy’s stellar mass. The distribution of $t_\mathrm{MWA}$ for the TITAN host galaxies is shown in Figure \ref{fig:age}, and spans a wide range from relatively young systems to galaxies dominated by old stellar populations, with a median value of \rev{7.1~Gyr}.

As expected, the inferred ages exhibit strong qualitative consistency with other host-galaxy properties. Galaxies with low sSFR and red rest-frame colours tend to occupy the older end of the distribution, while younger ages are associated with bluer, more actively star-forming systems. This agreement is important, as it indicates that the SED-fitting procedure is recovering physically meaningful variations in stellar population age despite the inherent model dependencies of age estimates.

Reliable age measurements are of particular interest in SN Ia cosmology because stellar population age is one of the most physically motivated quantities underlying many empirical host-galaxy trends. The commonly used host-mass correction is observationally convenient, but stellar mass is only a proxy for a network of correlated galaxy properties, including star-formation history, metallicity, dust content, and stellar age. Several recent studies have argued that host or progenitor age may be a more fundamental variable than stellar mass in driving residual environmental dependencies in SN Ia standardization, either through intrinsic luminosity differences between young and old progenitor populations or through changes in the effective dust law of SN environments \citep[e.g.,][]{Lee2022, Wiseman2023, Chung2025, Son2025}. This interpretation is further supported by environmental studies linking SN Ia demographics directly to the age of the surrounding stellar population. \rev{In particular, SNe~Ia in galaxy clusters, which are dominated by old, quiescent stellar populations, show a markedly different light curve stretch distribution from those in the field: \citet{Larison2024} found that SNe~Ia in the centres of nearby clusters are almost exclusively low-$x_1$ events, and \citet{Toy2023} found systematically lower stretch for cluster SNe~Ia in the Dark Energy Survey. These results suggest that stellar population age is closely tied to the SN~Ia progenitor population.}

This has motivated renewed debate over whether redshift evolution in the age distribution of SN Ia progenitors could bias cosmological inferences if not properly accounted for \citep[e.g.,][]{Chung2025, Son2025}. However, other analyses have argued that the size of this effect is small once standard host-mass corrections and realistic progenitor-age estimates are included \citep{Wiseman2026}. The TITAN host-galaxy age measurements therefore provide an essential foundation for directly testing these scenarios in a large, homogeneous low-redshift sample. \rev{In particular, \citet{MT26} combine SED-derived star-formation histories for the TITAN hosts with empirical delay-time distributions to infer progenitor-age distributions for all SNe~Ia in the sample, and assess the validity and cosmological impact of a potential progenitor-age bias. Their SFH modelling differs from that adopted here, and a comparison between the two is presented in their Appendix~A.}

\subsection{Local Properties}

The local photometry data are fitted with the same \texttt{Bagpipes} model as outlined in Section \ref{sec:sed_fitting} and Table \ref{tab:bagpipes_params}. \rev{When comparing local and global properties, we apply the quality cuts of \citet{Kelsey2023}: we require uncertainties of $\sigma_{g-z} < 1$\,mag in rest-frame colour and $\sigma_{M_*} < 1$\,dex in stellar mass for both the local and global fits, and that the global stellar mass exceeds the local one ($M_{*,\rm global} > M_{*,\rm local}$). These cuts reduce the comparison sample from 7329 to 7080 galaxies.}

\subsubsection{Stellar mass}

A comparison between the local and global stellar masses of the TITAN host galaxies is shown in the left-hand panel of Figure~\ref{fig:local_props}. As expected, the local stellar masses are systematically lower than the corresponding global values, reflecting the fact that the local apertures probe only a small fraction of the total host galaxy. Because each local measurement is integrated over a fixed 1.5~kpc aperture, the reported local mass effectively traces the stellar surface mass density at the SN explosion site rather than an independent mass scale. The strong correlation between local and global mass therefore indicates that galaxies with larger total stellar mass also tend, on average, to exhibit higher stellar surface densities at the SN location.

Despite this overall trend, the local masses exhibit substantial scatter at fixed global mass, highlighting the diversity of environments probed within galaxies of similar total stellar mass. Some SNe occur in relatively dense, centrally concentrated regions, while others are located in lower-density outer discs or diffuse stellar environments. This behaviour is closely tied to each SN's $d_\mathrm{DLR}$: systems with larger $d_\mathrm{DLR}$ values preferentially occupy the lower envelope of the local-mass distribution at fixed global mass, consistent with these SNe occurring in regions of lower stellar density relative to the host light profile.

\subsubsection{$g-z$ colour}

The relationship between local and global rest-frame $g-z$ colour is shown in the right-hand panel of Figure~\ref{fig:local_props}. \rev{The two are strongly correlated (Pearson $r = 0.69$), indicating that the stellar populations at the SN location broadly track those of the host galaxy as a whole, but with an intrinsic scatter of $\sim$0.14\,mag after accounting for the measurement uncertainties, such that the local environments of individual SNe Ia can differ significantly from the integrated properties of their hosts.}

\rev{The SN environments are also systematically redder than their hosts (median $\Delta(g-z) = +0.08$\,mag, with 70\% of SNe in locally redder regions), and this offset depends strongly on galactocentric distance: it decreases from $+0.18$\,mag for $d_{\rm DLR} < 0.5$, where 92\% of SNe lie in locally redder regions, to $\sim$0 for $2 < d_{\rm DLR} < 4$. This is consistent with the radial colour gradients of galaxies, whose central regions host older, redder stellar populations than their outskirts \citep[e.g.,][]{Perez2013, Sanchez2014, GonzalezDelgado2015, Belfiore2017, Goddard2017}, and with previous findings that the local environments of SNe Ia are redder than their hosts as a whole \citep{Roman2018, Kelsey2023, Kelsey2026}. For SNe within $d_{\rm DLR} < 2$, the correlation between local and global colour ($r = 0.87$) is comparable to that found by \citet{Kelsey2021} for 4\,kpc apertures ($r = 0.90$), despite our smaller 1.5\,kpc apertures; the weaker correlation for the full sample is driven by SNe in the faint outskirts of their hosts ($d_{\rm DLR} \geq 4$; $r = 0.44$), where the local photometry has low signal-to-noise.}

\rev{Such systematic differences between the inner and outer regions of galaxies may also affect SN Ia standardization: \citet{Toy2025} found that SNe Ia at larger $d_{\rm DLR}$ exhibit a smaller host-mass step than those near galaxy centres, which they attributed to differences in the underlying stellar populations between inner and outer regions. The strong dependence of the local--global colour offset on $d_{\rm DLR}$ found here supports this interpretation.}

\section{Conclusions}
\label{sec:conclusions}

In this paper we present a bespoke host-galaxy dataset for the TITAN SN~Ia sample, comprising 8378 spectroscopically confirmed SNe~Ia in the nearby Universe. The primary aims of this work are to identify the most likely host of each event, assemble the best available redshift information, measure consistent multi-band photometry, and derive global and local galaxy properties suitable for population studies and cosmology. A summary of the catalogue is presented in Appendix~\ref{app:catalogue}.

Our main conclusions are as follows.

\begin{enumerate}

    \item \textbf{Host association.} By combining multiple automatic host-association methodologies with manual inspection and vetting, we confidently assign host galaxies to 7856 SNe~Ia (93.8\% of the full TITAN sample). The remaining systems comprise 259 hostless SNe, 75 cases of ambiguous host association, and 188 events located at sufficiently large separations from nearby galaxies ($d_\mathrm{DLR}>12$) that no robust association can be made.

    \item \textbf{Redshift compilation.} We compile host-galaxy spectroscopic redshifts for 5522 SNe~Ia (65.9\% of the full TITAN sample). For the remainder, the best available redshifts are derived from SN spectra using \texttt{SNID-SAGE}. We also provide CMB-frame and peculiar-velocity-corrected redshifts based on \texttt{Manticore-Local} as part of the TITAN DR1 products.

    \item \textbf{Photometry and SED fitting.} Using \texttt{HostPhot}, we obtain global photometry spanning the FUV--MIR wavelength range, together with optical photometry measured within a fixed 1.5~kpc aperture centred on each SN position. These data are fitted using the \texttt{Bagpipes} SED-fitting framework, yielding successful global fits for \rev{7568} host galaxies and successful local fits for 7329 systems. For all successfully fitted hosts, we provide stellar masses and rest-frame $g-z$ colours as part of the TITAN DR1 products. For the subset of \rev{5827} galaxies with sufficient UV--MIR wavelength coverage \rev{and no sign of AGN contamination}, we additionally report global measurements of SFR, sSFR, dust attenuation ($A_V$), and mass-weighted stellar population age. We further demonstrate that these derived properties are robust to the absence of FIR constraints.

    \item \textbf{Sample properties.} The TITAN
    host-galaxy sample spans a wide and diverse range in stellar mass, colour, star-formation rate, dust attenuation, and stellar population age. The global mass distribution peaks at $\log_{10}(M_*/M_\odot) \sim 10$, with \rev{60\%} of SNe~Ia occurring in massive ($\log_{10}(M_*/M_\odot) > 10$) hosts and a median
    of $\log_{10}(M_*/M_\odot) = 10.22$. The TITAN host galaxies trace the familiar structure of the low-redshift galaxy population in the SFR--$M_*$ plane, with \rev{71\%} classified as star-forming and \rev{16\%} as quiescent. Local stellar masses and $g-z$ colours broadly track their global counterparts, but exhibit substantial scatter that reflects diversity in the SN progenitor environments within galaxies of similar integrated properties.

\end{enumerate}

Taken together, these results demonstrate that TITAN provides not only the largest homogeneous low-redshift SN~Ia sample assembled to date, but also an extensive and detailed host-galaxy catalogue designed to support the next generation of environmental and cosmological analyses. 

Forthcoming TITAN analyses will combine this catalogue with the SN observables presented by Murakami et al. (in prep.) to investigate the physical origin of the host-galaxy ``luminosity step'' and related environmental dependencies in SN Ia standardization. In particular, TITAN enables direct comparisons between stellar mass and more physically motivated quantities -- such as star-formation rate, dust attenuation, and stellar population age -- in order to determine whether the conventional host-mass correction is acting as a proxy for a more fundamental property of the SN environment.

Because host-galaxy effects now represent a major component of the systematic uncertainty budget in SN~Ia cosmology \citep{Vincenzi2024, Popovic2026}, building datasets of this kind is essential. TITAN provides an empirical framework for calibrating and interpreting those effects in the nearby Universe, and for preparing the environmental analyses that will be required for precision SN~Ia cosmology in the Rubin/LSST and \textit{Roman} era.

\section*{Acknowledgements}

\rev{The authors thank the anonymous referee for a thorough and comprehensive report that helped to materially improve the clarity, quality, and completeness of the data and results presented in this work.}

J.W.T. gratefully acknowledges support by the STFC through grant ST/Y509474/1. J.W.T. also thanks Matt Jarvis and Shruti Ramaiya for useful discussions.

MV is supported by UKRI Future Leaders Fellowship [UKRI2363].

SJS acknowledges funding from STFC Grants ST/Y001605/1, ST/X001253/1, a Royal Society Research Professorship and the Hintze Family Charitable Foundation. 

D.O.J. acknowledges support from NSF grants AST-2407632, AST-2429450, and AST-2510993, NASA grants 80NSSC24M0023 and 80NSSC24K0353, and HST/JWST grants HST-GO-17128.028 and JWST-GO-05324.031, awarded by the Space Telescope Science Institute (STScI), which is operated by the Association of Universities for Research in Astronomy, Inc., for NASA, under contract NAS5-26555. This work is also funded in part by the Gordon and Betty Moore Foundation through Grant GBMF13900 to D.O.J.

The authors thank the Hariri Institute at Boston University for their generous funding contributions to this project and for facilitating the space to hold a Summer 2025 collaboration meeting.

\section*{Data Availability}

The TITAN DR1 host-galaxy catalogue generated in this study\rev{, and a full table of photometric data,} will be made publicly available at \url{https://titan-snia.github.io} upon publication of this article. 

\bibliography{main}{}
\bibliographystyle{mnras}




\appendix

\section{Survey dependence of the directional light radius} \label{sec:app_larger_dlr}

In Section \ref{sec:dlr_cuts}, we found that the typical host association threshold of $d_{\mathrm{DLR}}\le4$ is overly restrictive for the TITAN sample, resulting in an unrealistically large fraction of orphan SNe Ia. In this appendix we justify our adopted host-association threshold of $d_{\mathrm{DLR}}\le12$ by drawing comparisons to the DES host galaxy sample \citep{Wiseman2020}, which adopts the same definition of $\mathrm{DLR}$ as this work \citep{Gupta2016}. 

\begin{figure}
    \centering
    \includegraphics[width=\columnwidth]{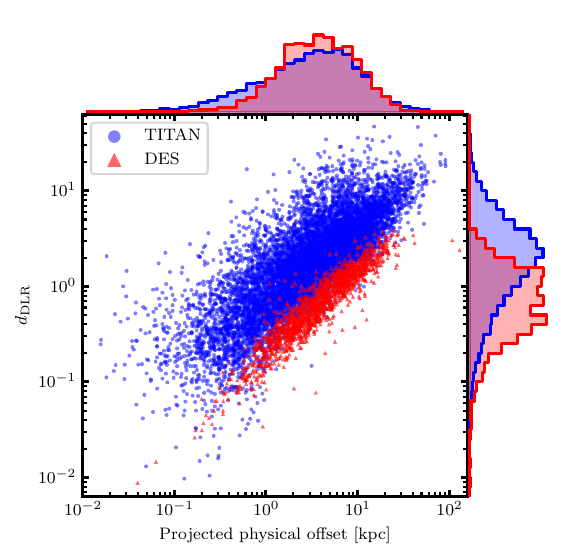}
    \caption{The relationship between DLR and physical offset between an SN and its host galaxy, shown for both the TITAN and DES samples. The marginal histograms show the corresponding 1D distributions.}
    \label{fig:titan_des_dlr_comparison}
\end{figure}

Using data from the ATLAS100 sample \citep{Srivastav2026}, we expect 90\% of SNe Ia to lie within a projected physical offset of $\sim11$~kpc of their host galaxy. As shown in Figure \ref{fig:titan_des_dlr_comparison}, we find that the data from both TITAN and DES are broadly consistent with this value, yet their $d_\mathrm{DLR}$ distributions differ significantly; a two-sided Kolmogorov--Smirnov test returns a $p$-value $<<0.05$. A $d_\mathrm{DLR}$ of 1 corresponds to a physical offset of $\sim2$~kpc in TITAN, compared with $\sim6$~kpc in DES; concordantly with this, to minimize the Wasserstein distance between the two distributions a multiplicative factor of 3 must be applied to the DES $d_\mathrm{DLR}$ distribution. We therefore scale the standard host-association threshold by the same factor, yielding our adopted host association threshold of $d_\mathrm{DLR}\leq12$. 

In an effort to explain why the correspondence between $d_\mathrm{DLR}$ and the projected physical offset differs so strongly between TITAN and DES, we present below a simplified analytic model that demonstrates why $d_{\mathrm{DLR}}$ is not invariant across imaging surveys even when defined consistently. The statistic depends on imaging depth and PSF, and is further modulated by cosmological surface-brightness dimming. 

\subsection{The directional light radius statistic}
\label{app:dlr_definition}

The semi-axes \texttt{A} and \texttt{B} used in Equation~\ref{eq:R_DLR} are derived from the flux-weighted second central moments of the host light profile, as returned by \texttt{sep} \citep{sep}. For a background-subtracted image with pixel intensities $I_{i}$ at positions $(x_{i}, y_{i})$, the flux-weighted centroid is
$\bar{x} = \sum_{i} I_{i}\,x_{i} \big/ \sum_{i} I_{i}$ and similarly
for $\bar{y}$, and the second central moments are
\begin{equation}
\begin{aligned}
M_{xx} &= \frac{\sum_{i} I_{i}\,(x_{i}-\bar{x})^{2}}{\sum_{i} I_{i}}, \\
M_{yy} &= \frac{\sum_{i} I_{i}\,(y_{i}-\bar{y})^{2}}{\sum_{i} I_{i}}, \\
M_{xy} &= \frac{\sum_{i} I_{i}\,(x_{i}-\bar{x})(y_{i}-\bar{y})}{\sum_{i} I_{i}}.
\end{aligned}
\label{eq:moments}
\end{equation}
where the sums run over all pixels in the isophotal area defined by the detection threshold. Diagonalizing the resulting moment matrix yields the semi-axes and position angle,
\begin{equation}
\begin{aligned}
\texttt{A}^{2} &= \frac{M_{xx}+M_{yy}}{2}
      + \sqrt{\left(\frac{M_{xx}-M_{yy}}{2}\right)^{2} + M_{xy}^{2}}, \\
\texttt{B}^{2} &= \frac{M_{xx}+M_{yy}}{2}
      - \sqrt{\left(\frac{M_{xx}-M_{yy}}{2}\right)^{2} + M_{xy}^{2}}.
\end{aligned}
\label{eq:AB}
\end{equation}
with $\texttt{A}\ge \texttt{B}$ by construction. 

The DLR of an elliptical aperture with semi-major axis \texttt{A}, semi-minor axis \texttt{B}, and SN--host position angle \texttt{THETA} is defined in Equation \ref{eq:R_DLR}, with the translation to $d_\mathrm{DLR}$ defined in Equation \ref{eq:ddlr}. Because both $d_\perp$ and $R_\mathrm{DLR}$ scale with redshift as the inverse angular diameter distance $1/D_{A}(z)$, the purely geometric redshift dependence cancels in Equation \ref{eq:ddlr}, hence motivating the use of $d_{\rm DLR}$ as a distance-independent association statistic.

\subsection{Survey-dependent measurement of the second-moment ellipse}
\label{app:dlr_survey_dependence}

The cancellation of $D_{A}$ in Equation~\ref{eq:ddlr} requires that \texttt{A} and \texttt{B} be measured consistently for the same physical galaxy across surveys. This condition is not met when the two surveys have different imaging depths or PSFs, because the second-moment integral is taken over the isophotal area defined by the local detection threshold. For a host profile of intrinsic rest-frame surface brightness $\mu_{\rm rest}(r)$, the truncation radius $R_{t}$ in a fixed observing band is set by
\begin{equation}
\mu_{\rm rest}(R_{t}) \;=\; \mu_{\rm lim,rest}(z)
   \;\equiv\; \mu_{\rm lim,obs} - 7.5\log_{10}(1+z) - K(z),
\label{eq:mu_lim_rest}
\end{equation}
where we account both for cosmological surface brightness dimming \citep{Tolman1930, Tolman1934} and the $K$-correction, $K(z)$.

The majority of TITAN $d_\mathrm{DLR}$ values are derived from Pan-STARRS1 $3\pi$ stacks, which reach $r$-band 5$\sigma$ point-source depths of $\sim23.2$ \citep{PanSTARRS}, which corresponds at the PSF-area (equation~2 of \citealt{Wiseman2020}) convention to a surface brightness limit of $\mu_{\rm lim,obs}^{\rm PS1}\simeq24.6$ \,mag\,arcsec$^{-2}$. The DES SN-field deep coadds of \citet{Wiseman2020} reach $\mu_{\rm lim,obs}^{\rm DES}\simeq27.2$\,mag\,arcsec$^{-2}$ at the same convention -- a difference of $\sim2.6$\,mag in observed depth, set primarily by the much larger total integration time of the DES coadds.

The measured second-moment ellipse is the convolution of the truncated intrinsic profile with the PSF, so the per-axis variances approximately add:
\begin{equation}
\begin{aligned}
\texttt{A}_{\rm phys}^{2} &= \sigma_{\rm gal,phys}^{2}(z) + \sigma_{\rm PSF,phys}^{2}(z), \\
\texttt{B}_{\rm phys}^{2} &= q^{2}\sigma_{\rm gal,phys}^{2}(z) + \sigma_{\rm PSF,phys}^{2}(z).
\end{aligned}
\label{eq:AB_obs}
\end{equation}
where $q\equiv b/a$ is the intrinsic isophotal axis ratio of the galaxy and $\sigma_{\rm PSF,phys}(z) = (\mathrm{FWHM}/2.355)\,\mathfrak{s}(z)$ with $\mathfrak{s}(z)$ the angular-diameter scale in kpc/arcsec; we take $\mathrm{FWHM}_\mathrm{PS1, r}=1.19^{\prime\prime}$ \citep{Magnier2020} and $\mathrm{FWHM}_\mathrm{DECam, r}=0.9^{\prime\prime}$ \citep{Wiseman2020}. Note that a fixed-arcsec PSF subtends a larger physical scale at higher redshift, so $\sigma_{\rm PSF,phys}$ inflates the DES measurement more
than the TITAN one.

\subsection{Quantitative model}
\label{app:dlr_model}

We model the host as a S\'ersic \citep{Sersic1963, Sersic1968} profile of index $n$, effective radius $R_{e}$, and surface brightness at $R_{e}$ of $\mu_{e}$:
\begin{equation}
\mu_{\rm rest}(m) = \mu_{e} + \frac{2.5\,b_{n}}{\ln 10}
   \Bigl[\bigl(m/R_{e}\bigr)^{1/n} - 1\Bigr],
\label{eq:sersic_mu}
\end{equation}
with elliptical radius $m^{2} = x^{2} + (y/q)^{2}$ and $b_{n}$ the Ciotti--Bertin coefficient \citep{CiottiBertin1999}, $b_{n} \approx 2n - 1/3 + 4/(405n)$. The flux-weighted second moment per axis, truncated at $R_{\rm t}$, is
\begin{equation}
\sigma_{\rm gal,phys}^{2}(z) \;=\;
   \tfrac{1}{2}\,R_{e}^{2}\,b_{n}^{-2n}\,
   \frac{\gamma(4n,\,t_\mathrm{lim}(z))}{\gamma(2n,\,t_\mathrm{lim}(z))},
\label{eq:sigma_gal}
\end{equation}
where $\gamma(s,t) = \int_{0}^{t} u^{s-1}e^{-u}\,du$ is the lower incomplete gamma function and
\begin{equation}
t_\mathrm{lim}(z) \;=\; b_{n} + \frac{\ln 10}{2.5}\bigl[\mu_{\rm lim,rest}(z) - \mu_{e}\bigr]
       \;=\; b_{n} + 0.921\,\bigl[\mu_{\rm lim,rest}(z) - \mu_{e}\bigr]
\label{eq:y_arg}
\end{equation}
is the dimensionless truncation argument obtained by setting $\mu_{\rm rest}(R_{t})=\mu_{\rm lim,rest}$ and substituting $u = b_{n}(m/R_{e})^{1/n}$ in the moment integrals. In the limit $t_\mathrm{lim}\to\infty$ (infinite depth), $\sigma_{\rm gal,phys}\to R_{e}\,b_{n}^{-n} \sqrt{\Gamma(4n)/(2\Gamma(2n))}$, giving $\sigma_{\rm gal,phys}\simeq
1.03\,R_{e}$ for $n=1$ but $\simeq 3.29\,R_{e}$ for $n=4$: the second moment of a de Vaucouleurs profile is dominated by its outer wings, making it especially sensitive to the survey's $\mu_{\rm lim}$.

Since the SN--host offsets $d_\perp$ are in agreement between both surveys, the ratio of the dimensionless statistics is
\begin{equation}
\frac{d_{\rm DLR}^{\rm TITAN}}{d_{\rm DLR}^{\rm DES}} \;=\;
   \frac{\mathrm{DLR}^{\rm DES}_\mathrm{phys}(\theta;z_{\rm DES})}
        {\mathrm{DLR}^{\rm TITAN}_\mathrm{phys}(\theta;z_{\rm TITAN})},
\label{eq:dDLR_ratio}
\end{equation}
where $\mathrm{DLR}_\mathrm{phys}$ is computed using the variances defined in Equation \ref{eq:AB_obs}.

\subsection{Predicted $d_\mathrm{DLR}$ shift between TITAN and DES}
\label{app:dlr_numerics}

\begin{figure*}
    \centering
    \includegraphics[width=\linewidth]{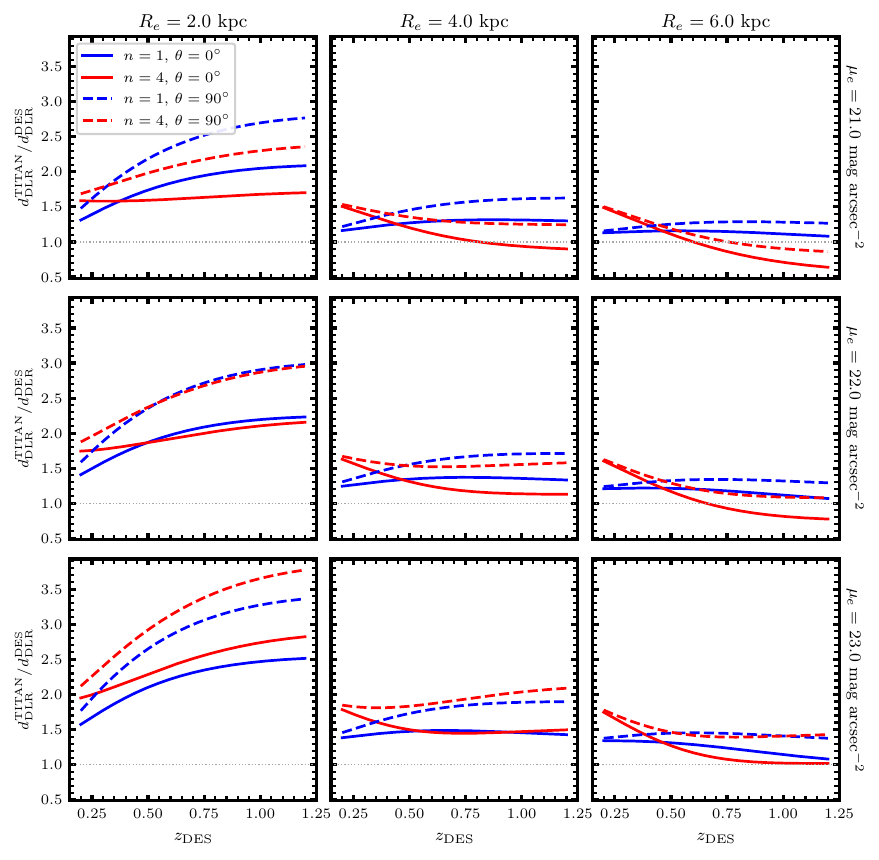}
    \caption{The ratio $d_{\rm DLR}^{\rm TITAN}/d_{\rm DLR}^{\rm DES}$ as a function of $z_{\rm DES}$ at fixed $z_{\rm TITAN}=0.05$, for disc ($n=1$, blue) and elliptical ($n=4$, red) S\'ersic profiles with axis ratio $q=0.7$. Each column varies the effective radius ($R_e = 2$, $4$, $6$ kpc) and each row varies the rest-frame surface brightness at the effective radius ($\mu_e = 21$, $22$, $23$ mag arcsec$^{-2}$). Solid and dashed lines show the ratio for a supernova located along the semi-major ($\theta=0^\circ$) and semi-minor ($\theta=90^\circ$) axis respectively; the spread between them reflects the intrinsic ellipticity of the profile. K-corrections appropriate to PS1-$r$ (TITAN) and DECam-$r$ (DES) are applied.}
    \label{fig:dlr_ratio_comparison}
\end{figure*}

We evaluate Equation~\ref{eq:dDLR_ratio} across a representative grid of galaxy parameters in Figure \ref{fig:dlr_ratio_comparison}. $K$-corrections for the PS1-$r$ (TITAN) and DECam-$r$ (DES) passbands are computed from FSPS \citep{Conroy2009, ForemanMackey2014} spectral templates: an SSP ($t_{\rm age} = 10$~Gyr, solar metallicity) for $n=4$ galaxies and a delayed exponential SFH ($\tau = 3$~Gyr, $t_{\rm age} = 7$~Gyr) for $n=1$ galaxies, convolved with the PS1-$r$ and DECam-$r$ filter curves.

The ratio exceeds unity across most of the parameter space explored, confirming that TITAN systematically assigns larger normalized offsets to the same SN--host pair relative to DES. For compact disc galaxies ($R_e = 2$~kpc, $n = 1$), the PSF inflation effect dominates over the entire redshift range modelled; as $z_{\rm DES}$ increases the DES PSF in physical kpc grows, producing a progressively larger apparent semi-axis. For large discs ($R_e = 6$~kpc), however, the galaxy already dominates over the PSF at all redshifts, and surface brightness dimming steadily erodes the detected outer disc, partially offsetting the PSF inflation. For large elliptical galaxies, the redder early-type SED incurs a substantially larger $K$-correction than for late-type galaxies, progressively suppressing the DES rest-frame limiting isophote with increasing redshift and eroding its depth advantage, sometimes resulting in $d_\mathrm{DLR}^{\mathrm{DES}} > d_\mathrm{DLR}^{\mathrm{TITAN}}$. 

In all cases, a fainter $\mu_{e}$ reduces $t_\mathrm{lim}$, resulting in steeper growth of the incomplete gamma function which translates to increased sensitivity to additional depth. Therefore, the fixed 2.6~mag arcsec$^{-2}$ depth difference between TITAN and DES results in a larger fractional gain in detected extent for faint galaxies, and thus an increase in the $d_\mathrm{DLR}$ ratio. Furthermore, SNe located along the semi-minor axis always incur a larger intersurvey discrepancy than those along the semi-major axis, regardless of morphological type, with the difference increasing with redshift. This is a direct effect of PSF circularization, to which the semi-minor axis is significantly more susceptible than the semi-major axis, since the relative impact of a growing PSF is larger for shorter length scales. For a randomly oriented SN, the relevant DLR averages over $\theta$ and so the population-level shift will lie between the two bracketing curves.

\subsection{Comparison with the observed distribution and adopted threshold}
\label{app:dlr_choice}

The model above isolates the depth-, PSF-, and dimming-driven component of the cross-survey shift in $d_\mathrm{DLR}$. Empirically, the TITAN and DES $d_\mathrm{DLR}$ distributions differ by a factor of $\sim3$ in their medians, motivating our adopted association threshold of $d_\mathrm{DLR}\leq12$ for the TITAN sample. Our simplified analytic model demonstrates that TITAN $d_\mathrm{DLR}$ values are typically expected to be larger than those of DES, and that a $d_\mathrm{DLR}$ ratio of 3 is indeed feasible on a host-by-host basis, within the compact, low-surface-brightness, and minor-axis regions of parameter space; for well-resolved bright hosts the predicted shift is smaller, so the exact population median will depend on the joint distribution of host properties.

Several simplifications in the model likely cause it to underestimate the true inter-survey DLR ratio. For example, real galaxies are not described by a single smooth S\'ersic component. Bulge--disc hosts in particular have a bright bulge contributing most of the central surface brightness but only a modest fraction of the second moment, and a much fainter disc extension whose flux dominates the root-mean-square despite sitting close to the surface-brightness limit. In shallow imaging, the disc component sinks below the detection isophote while the bulge remains bright, and the recovered second moments collapse onto a bulge-dominated measurement with a much smaller \texttt{A} than the single-S\'ersic prediction. The model has no degree of freedom representing this and so understates the depth-induced shrinkage for low-redshift early-type spirals -- which constitute a substantial fraction of low-$z$ hosts in TITAN. The PSF treatment introduces two further sources of underestimation. Although adding the PSF and galaxy variances in quadrature is exact for second moments regardless of profile shape, the Gaussian PSF assumption is not: real ground-based PSFs have broader wings than a Gaussian of the same FWHM, so $\sigma_{\rm PSF} > {\rm FWHM}/2.355$ and the model underestimates the PSF contribution to the observed size.  A more subtle issue is the order of operations: the model truncates the intrinsic profile at $\mu_{\rm lim,rest}$ before applying the PSF, whereas in reality PSF convolution redistributes flux outward first, pushing the detectable outer isophote of the PSF-convolved image to larger radii than the truncation radius of the unconvolved profile. Both effects are likely more significant for DES, where the physical scale of the PSF is larger at $z_{\rm DES} >> 0.05$, and both act to increase the DES DLR relative to the model prediction. Finally, the galaxy size--surface brightness relation \citep[the Kormendy relation;][]{Kormendy1977} evolves with redshift: at fixed stellar mass, galaxies are smaller and brighter in their central regions at earlier cosmic epochs \citep{vanderWel2014}. The DES sample therefore contains a higher fraction of compact, high-surface-brightness hosts than TITAN, whose PSF-dominated light profiles result in artificially smaller $d_\mathrm{DLR}$ values. These, and other, effects suggest the ratios shown in Figure~\ref{fig:dlr_ratio_comparison} are conservative lower bounds, and that the true inter-survey bias in $d_{\rm DLR}$ for the most problematic hosts could plausibly reach or exceed a factor of~3.

\section{\rev{Impact of SN-based redshifts on derived host properties}}
\label{app:snz}
 
\rev{The redshift of each host is fixed in the SED fitting (Section~\ref{sec:fiducial_sed_fitting}), so any error in the adopted redshift propagates into the derived properties. To quantify the impact of adopting an SN-based redshift, we refit the global photometry of the 2236 hosts in the TITAN sample that were classified from SEDM spectra and also have a host-galaxy spectroscopic redshift, fixing $z$ to the \texttt{SNID-SAGE} value rather than the host-galaxy redshift, with all other aspects of the fit unchanged. SEDM is the instrument responsible for the largest fraction of TITAN classifications, and yields the least precise SN-based redshifts (Section~\ref{sec:sn_z}), so this sample provides a conservative estimate of the impact on hosts relying on SN-based redshifts. For these hosts, $\Delta z = z_{\rm SN} - z_{\rm host, spec}$ has a median (nMAD) of $+0.003$ (0.004); 291 hosts have $|\Delta z| > 0.01$, 58 have $|\Delta z| > 0.02$, and 5 have $|\Delta z| > 0.05$. For the 171 hosts with $|\Delta z| < 0.0005$, the two sets of fits agree to within the scatter between repeated fits of the same photometry, confirming that the differences described below arise from the change in redshift alone.}
 
\begin{figure*}
\centering
\includegraphics[width=\textwidth]{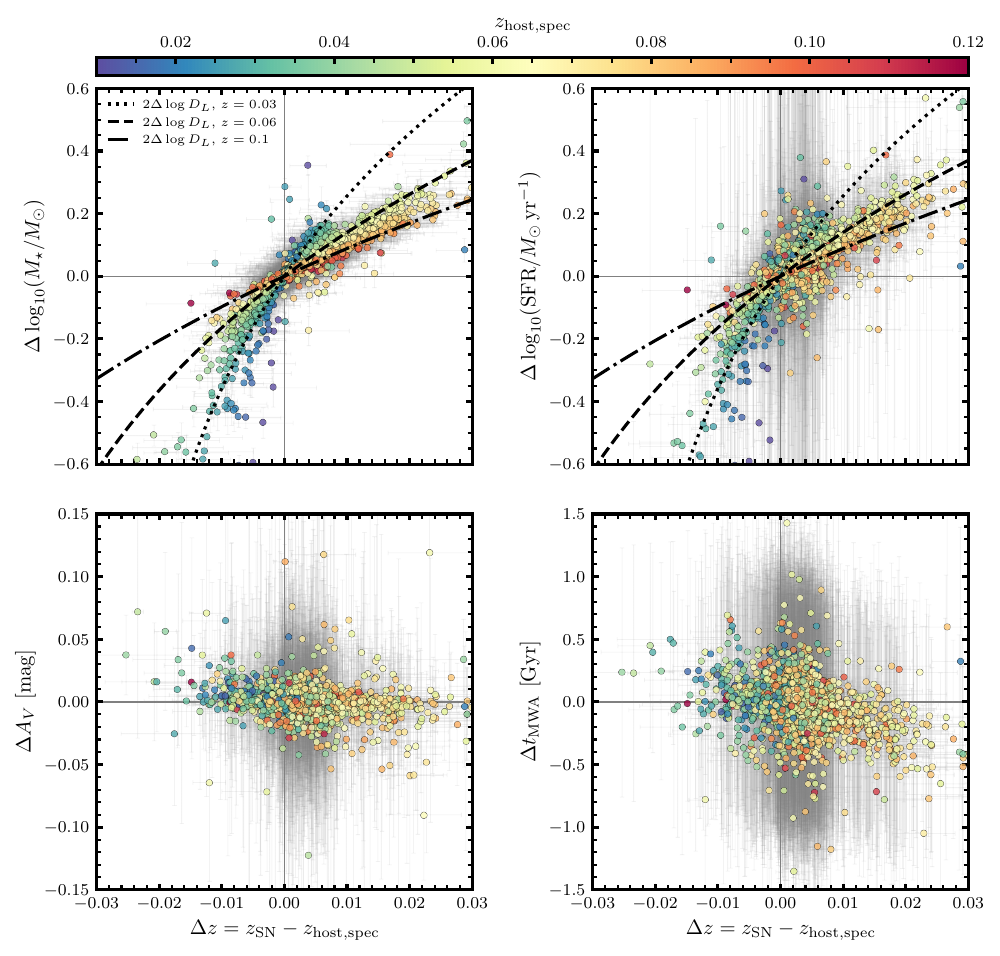}
\caption{\rev{Change in the inferred global host properties when the redshift used in the SED fit is fixed to the SN-based \texttt{SNID-SAGE} value rather than the host-galaxy spectroscopic redshift, as a function of $\Delta z = z_{\rm SN} - z_{\rm host}$, for the 2236 hosts in the TITAN sample with SEDM classifications and host-galaxy redshifts. Top left to bottom right: stellar mass, SFR, $A_V$, and mass-weighted age, each shown as the SN-redshift value minus the fiducial value. Points are coloured by host redshift. Black curves show the change expected from the change in luminosity distance alone, $2\,\Delta\log_{10}D_L$, at $z = 0.03$, 0.06 and 0.10. Error bars show the 1$\sigma$ statistical uncertainties.}}
\label{fig:snz_hostprops}
\end{figure*}

\rev{The results are shown in Figure~\ref{fig:snz_hostprops}. The change in stellar mass closely follows the change in luminosity distance, $\Delta\log_{10}M_* \simeq 2\,\Delta\log_{10}D_L$; this is expected as the SED fit infers luminosities from the observed fluxes via $L = 4\pi D_L^2 F$, while the mass-to-light ratio is constrained by the shape of the SED, which is independent of distance. Since $D_L \propto z$ at low redshift, $\Delta\log_{10}M_* \approx 0.87\,\Delta z/z$, so that a given redshift error has a larger impact on nearer hosts, as seen in the dependence on $z_{\rm host}$ in Figure~\ref{fig:snz_hostprops}. Across the sample, the median change in $\log_{10}M_*$ is $+0.04$\,dex, reflecting the small positive offset in the SEDM redshifts, with an nMAD of 0.06\,dex, comparable to the median statistical uncertainty of 0.04\,dex; 7.7 (3.3)\% of hosts shift by more than 0.2 (0.3)\,dex. The SFR scales with distance in the same way (median change $+0.04$\,dex, nMAD 0.07\,dex).}
 
\rev{In contrast, the distance-independent properties are essentially unaffected: the changes in sSFR, $A_V$ and rest-frame $g-z$ colour (nMAD of 0.025\,dex, 0.005\,mag and 0.008\,mag, respectively) are comparable to the scatter between repeated fits, and far smaller than their statistical uncertainties. The mass-weighted age shows a weak trend with $\Delta z$, decreasing by $\sim$0.2--0.3\,Gyr for $\Delta z \sim +0.02$. This arises because a higher adopted redshift implies a younger Universe at the epoch of the host, which limits the maximum age of its stellar population in the fit; at $z \sim 0.05$, the age of the Universe decreases by $\sim$0.3\,Gyr for $\Delta z = +0.025$. This effect is small compared with the typical statistical uncertainty on $t_{\rm MWA}$ ($\sim$0.9\,Gyr).}
 
\rev{We conclude that, for hosts relying on SN-based redshifts, stellar masses and SFRs carry an additional uncertainty of $\sim$0.06\,dex, whereas sSFR, $A_V$, $t_{\rm MWA}$ and rest-frame colours are robust to the choice of redshift. We propagate this uncertainty into the catalogue by adding $0.87\,\sigma_z/z$, where $\sigma_z$ is the uncertainty on the \texttt{SNID-SAGE} redshift, in quadrature to the uncertainties on $\log_{10}M_*$ and $\log_{10}{\rm SFR}$ for hosts relying on an SN-derived redshift.}
 
\section{Excluding multiple wavelength regimes \& individual photometric bands}
\label{app:excl_phot}
 
\begin{figure*}
\centering
\includegraphics[width=\textwidth]{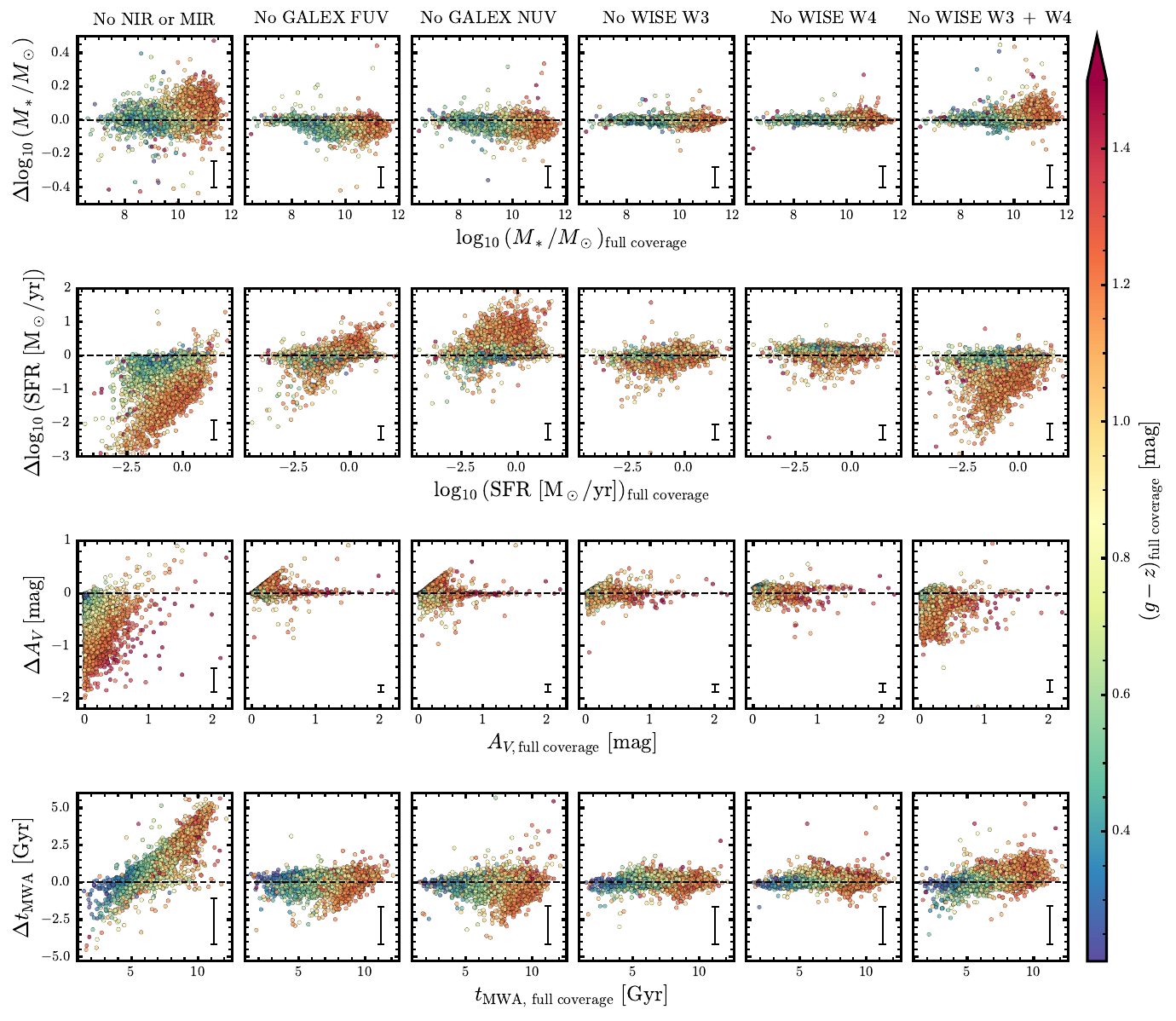}
\caption{\revtwo{As Figure~\ref{fig:effect_of_missing_photometry}, but for fits excluding combinations of wavelength regimes or individual bands. From left to right, the columns show fits excluding the NIR and MIR together, GALEX FUV (retaining the NUV), GALEX NUV (retaining the FUV), WISE W3 (retaining W4), WISE W4 (retaining W3), and both W3 and W4 (retaining W1 and W2). From top to bottom, the rows show stellar mass ($M_*$), SFR, dust attenuation ($A_V$) and mass-weighted age ($t_{\rm MWA}$). Points are coloured by the full-coverage rest-frame $g - z$ colour, error bars show the median statistical uncertainty of the difference, and horizontal dashed lines mark zero offset.}}
\label{fig:excl_phot_app}
\end{figure*}

\revtwo{Figure~\ref{fig:effect_of_missing_photometry} shows the effect on the global host properties of excluding an entire wavelength regime from the SED fits. Here we present the remaining tests, which exclude combinations of regimes or individual bands within a regime in Figure~\ref{fig:excl_phot_app}. These refine the results of Section~\ref{sec:excluding_phot} and motivate the coverage requirement adopted there: a measurement in both GALEX bands and in at least one of WISE W3 or W4. As in Figure~\ref{fig:effect_of_missing_photometry}, all offsets are relative to fits with full coverage, $\Delta X = X_{\rm full} - X_{\rm restricted}$, and are shown for the hosts with full coverage.}
 
\section{Catalogue Summary} \label{app:catalogue}

In Table \ref{tab:catalogue_columns} we provide a summary of all columns comprising the TITAN host galaxy catalogue. \rev{Alongside this table, we will also release a table containing all host galaxy photometry. These tables complement the corresponding light curve parameter table which will be released upon the publication of Murakami et al. (in prep.).}

\begin{table*}
\centering
\caption{Summary of TITAN host catalogue columns.}
\label{tab:catalogue_columns}
\begin{tabular}{lll}
\hline
\textbf{Column} & \textbf{Unit} & \textbf{Description} \\
\hline

objname & -- & Supernova name \\
\rev{sample\_membership} & \rev{--} & \rev{Whether the SN belongs to the bronze, silver or gold sample} \\
sn\_ra & deg & Supernova right ascension \\
sn\_dec & deg & Supernova declination \\
host\_ra & deg & Host-galaxy right ascension \\
host\_dec & deg & Host-galaxy declination \\
host\_z(\_err) & -- & (Uncertainty on) host-galaxy redshift \\
host\_z\_source & -- & Source of host-galaxy redshift \\
\rev{pv\_manticore(\_err)} & km\,s$^{-1}$ & (Uncertainty on) \texttt{Manticore-Local} peculiar velocity correction \\
\rev{pv\_2mpp(\_err)} & \rev{km\,s$^{-1}$} & \rev{(Uncertainty on) 2M++ peculiar velocity correction} \\
\rev{z\_hel(\_err)} & \rev{--} & \rev{(Uncertainty on) heliocentric redshift$^{a}$} \\
\rev{z\_cmb(\_err)} & \rev{--} & \rev{(Uncertainty on) CMB-frame redshift} \\
\rev{z\_hd(\_err)} & \rev{--} & \rev{(Uncertainty on) Hubble diagram redshift$^{b}$} \\
host\_catalogue & -- & catalogue of origin of the host galaxy \\
host\_name & -- & Name of host galaxy within the catalogue of origin \\
sn\_host\_sep\_arcsec & arcsec & Angular separation between SN and host centre \\
sn\_host\_sep\_kpc & kpc & Physical projected separation between SN and host centre \\
d\_dlr & -- & Directional light radius host-association metric \\
global\_stellar\_mass\_\{16,50,84\} & log$_{10}(M_\ast/M_\odot)$ & (16th, 50th, 84th) percentile global stellar mass \\
global\_sfr\_\{16,50,84\} & log$_{10}(M_\odot\,\mathrm{yr}^{-1})$ & (16th, 50th, 84th) percentile global star-formation rate \\
global\_ssfr\_\{16,50,84\} & log$_{10}(\mathrm{yr}^{-1})$ & (16th, 50th, 84th) percentile global specific star-formation rate \\
global\_Av\_\{16,50,84\} & mag & (16th, 50th, 84th) percentile global dust attenuation $A_V$ \\
global\_mass\_weighted\_age\_\{16,50,84\} & Gyr & (16th, 50th, 84th) percentile global mass-weighted stellar age \\
\rev{global\_gz\_colour\_\{16,50,84\}} & \rev{AB mag} & \rev{(16th, 50th, 84th) percentile rest-frame global ($g-z$) colour} \\
local\_stellar\_mass\_\{16,50,84\} & log$_{10}(M_\ast/M_\odot)$ & (16th, 50th, 84th) percentile stellar mass within 1.5~kpc of SN position \\
\rev{local\_gz\_colour\_\{16,50,84\}} & \rev{AB mag} & \rev{(16th, 50th, 84th) percentile rest-frame local ($g-z$) colour} \\

\hline
\end{tabular}
\vspace{2mm}
\begin{minipage}{\textwidth}
\footnotesize
\rev{$^{a}$\,The heliocentric redshift is defined as the best available redshift: it is the spectroscopic host redshift, if available, and the redshift derived from the SN spectrum otherwise.} \\
\rev{$^{b}$\,The Hubble diagram redshift in the catalogue is derived using the \texttt{Manticore-Local} peculiar velocity correction.}
\end{minipage}

\end{table*}


\bsp	
\label{lastpage}
\end{document}